\documentclass{ieeeaccess}
\usepackage{cite}
\usepackage{amsmath,amssymb,amsfonts}
\usepackage{algorithmic}
\usepackage{graphicx}
\usepackage{textcomp}
\usepackage{multirow}
\usepackage{multicol}
\usepackage{colortbl}
\usepackage{hhline}
\usepackage{color,soul}
\usepackage[table]{xcolor} 
\usepackage[linesnumbered,ruled,vlined]{algorithm2e}
\SetAlFnt{\small}
 \usepackage{url}

\SetCommentSty{mycommfont}

\SetKwInput{KwInput}{Input}                
\SetKwInput{KwOutput}{Output}              

\def\BibTeX{{\rm B\kern-.05em{\sc i\kern-.025em b}\kern-.08em
    T\kern-.1667em\lower.7ex\hbox{E}\kern-.125emX}}

\begin{document}
\doi{10.1109/ACCESS.2020.3029339}

\title{A Survey on Split Manufacturing: Attacks, Defenses, and Challenges}
\author{\uppercase{Tiago D. Perez},
\uppercase{Samuel Pagliarini}}
\address[]{Tallinn University of Technology (TalTech) \\
Department of Computer Systems \\
Centre for Hardware Security \\
Tallinn, Estonia\\ (e-mails: \{tiago.perez,samuel.pagliarini\}@taltech.ee)}
\tfootnote{This work was supported by the European Union through the European Social Fund in the context of the project ``ICT programme''.}

\newcommand{\etal}{\textit{et al.}}

\markboth
{Perez \headeretal: A Survey on Split Manufacturing: Attacks, Defenses, and Challenges}
{Perez \headeretal: A Survey on Split Manufacturing: Attacks, Defenses, and Challenges}

\corresp{Corresponding author: T. D. Perez (e-mail: tiago.perez@taltech.ee).}

\begin{abstract}

In today's integrated circuit (IC) ecosystem, owning a foundry is not economically viable, and therefore most IC design houses are now working under a fabless business model. In order to overcome security concerns associated with the outsorcing of IC fabrication, the Split Manufacturing technique was proposed. In Split Manufacturing, the Front End of Line (FEOL) layers (transistors and lower metal layers) are fabricated at an untrusted high-end foundry, while the Back End of Line (BEOL) layers (higher metal layers) are manufactured at a trusted low-end foundry. This approach hides the BEOL connections from the untrusted foundry, thus preventing overproduction and piracy threats. However, many works demonstrate that BEOL connections can be derived by exploiting layout characteristics that are introduced by heuristics employed in typical floorplanning, placement, and routing algorithms. Since straightforward Split Manufacturing may not afford a desirable security level, many authors propose defense techniques to be used along with Split Manufacturing. In our survey, we present a detailed overview of the technique, the many types of attacks towards Split Manufacturing, as well as possible defense techniques described in the literature. For the attacks, we present a concise discussion on the different threat models and assumptions, while for the defenses we classify the studies into three categories: proximity perturbation, wire lifting, and layout obfuscation. The main outcome of our survey is to highlight the discrepancy between many studies -- some claim netlists can be reconstructed with near perfect precision, while others claim marginal success in retrieving BEOL connections. Finally, we also discuss future trends and challenges inherent to Split Manufacturing, including the fundamental difficulty of evaluating the efficiency of the technique.
 
\end{abstract}


\titlepgskip=-15pt

\maketitle

\section{Introduction}
\label{sec:introduction}
Counterfeiting and intellectual property (IP) infringement are growing problems in several industries, including the electronics sector. In Europe, for instance, seizures of counterfeit electronics products increased by almost 30\% when comparing the 2014-2016 to the 2011-2013 period \cite{EUIPO19}. Legitimate electronics companies reported about \$100 billion in sales losses every year because of counterfeiting \cite{BogusIEEE}.
 
 \Figure[t!](topskip=0pt, botskip=0pt, midskip=0pt)[width=1\linewidth]{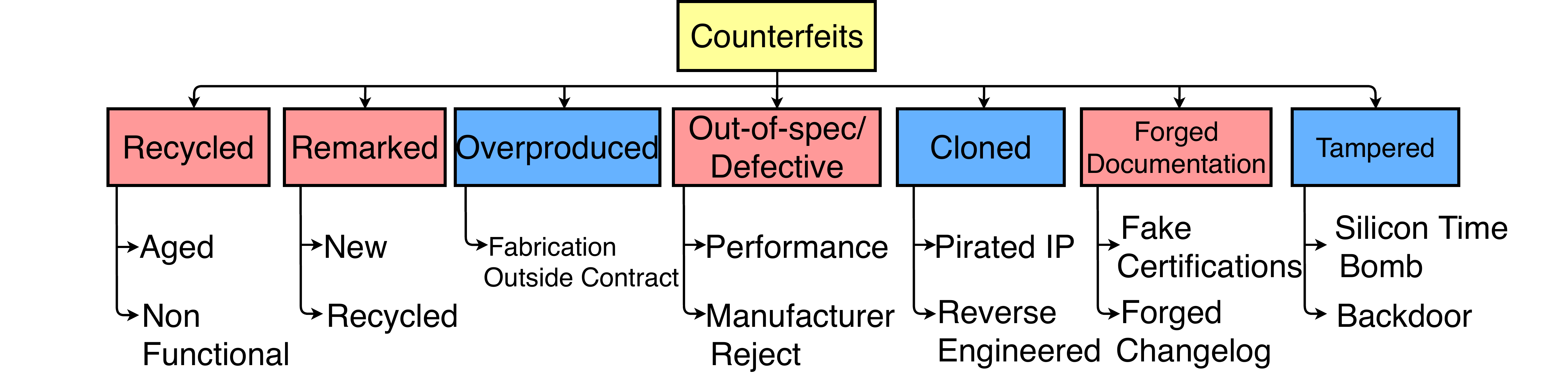}
{Taxonomy of counterfeit electronics (adapted from \cite{Guin2014}). \label{fig:cf_tax}}

As electronic systems are being increasingly deployed in critical infrastructure, counterfeit and maliciously modified integrated circuits (ICs) have become a major concern. The globalized nature of the IC supply chain contributes to the problem as we lack the means to assess the trustworthiness of the design and fabrication of ICs. It is conceivable -- if not likely -- that a fault in a low-quality counterfeit IC (or even a maliciously modified IC) will effectively disrupt critical infrastructure with grave consequences. Therefore, hardware security has gained more attention in the past decades, emerging as a relevant research topic.

As the IC supply chain has become more globalized, ensuring the integrity and trustworthiness of ICs becomes more challenging \cite{Guin2014}. When a modern IC is conceived, the probability that all involved parties are trusted is, in practice, close to zero. The process of conceiving an IC can be broken down into three major steps: design, manufacturing, and validation. \textit{Designing} an IC involves arranging blocks and their interconnections. Some blocks are in-house developed, while some are third-party IPs. Finally, a layout is generated by instantiating libraries that might also be in-house developed or provided by third parties. The resulting layout is then sent to a foundry for \textit{manufacturing}. The process of \textit{validation} requires test for physical defects as well as verification of packaged parts for correct functionality. Both test and packaging facilities may be untrusted, as these efforts are often offshored. Thus, in order to produce an IC, sensitive information almost inevitably is exposed to untrusted parties. Today's reality is that ICs are vulnerable to many hardware-based threats, including insertion of hardware trojans, IP piracy, IC overbuilding, reverse engineering, side-channel attacks, and counterfeiting. These threats are discussed in details in \cite{Rostami2014}. 

Hardware trojans, in particular, are malicious modifications to an IC, where attackers insert circuitry (or modify the existing logic) for their own malicious purposes. This type of attack is (typically) mounted during manufacturing, as the foundry holds the entire layout and can easily identify critical locations for trojan insertion. Third-party IPs can also contain trojans/backdoors that may contain hidden functionalities, and which can be used to access restricted parts of the design and/or expose data that would otherwise be unknown to the adversary.

IP piracy and IC overbuilding are, essentially, illegal ownership claims of different degrees. As said before, during designing an IC, third-party and in-house developed IPs are utilized. The untrusted foundry (or a rogue employee of it) can copy one of those IPs without the owner's authorization. Similarly, malicious foundries can manufacture a surplus of ICs (overbuilding) without the owner's knowledge, and sell these parts in the grey market.

Reversing engineering of ICs has been extensively demonstrated in the specialized literature \cite{Torrance2011}. An attacker can identify the technology node and underlying components (memory, analog, and standard cells), from which a gate-level netlist can be extracted and even a high-level abstraction can be inferred \cite{reverse}. Reverse engineering can be effortlessly executed during manufacturing, as the foundry holds the entire layout and most likely promptly recognizes some of the IP as well. After fabrication, -- when ICs are already packaged and deployed -- reverse engineering is more laborious but can still be executed by a knowledgeable adversary.

According to \cite{Guin2014}, counterfeit components are classified into seven distinct categories, as illustrated in Figure~\ref{fig:cf_tax}. Recycled, remarked, out-of-spec/defective, and forged documentation are intrinsic after-market problems, where products are offered by parties other than the original component manufacturer or authorized vendors. These cases are highlighted in red. On the other hand, overproducing, cloning, and tampering are problems faced during the designing and/or fabrication of ICs. These cases are highlighted in blue. For this reason, in this paper, we will focus on these threats. It is important to realize that these threats could be avoided if a trusted fabrication scheme was in place. However, the escalating cost and complexity of semiconductor manufacturing on advanced technologies made owning an advanced foundry unfeasible for design companies, which now have the tendency to adopt the fabless business model \cite{NewsAnySillion}. Consequentially, outsourcing of the manufacturing exposes their entire layout to untrusted foundries, leaving their designs vulnerable to malicious attacks.
  
While many \textit{ad hoc} techniques have been proposed to individually combat these threats, very few solutions directly address the lack of trust in the fabrication process. Split Manufacturing stands out from other techniques as it promotes a hybrid solution between trusted and untrusted fabrication. The technique was first pitched to DARPA circa 2006 in a white paper authored by Carnegie Mellon and Stanford universities. Later, it was picked by IARPA which then launched the Trusted IC program \cite{IARPA1} that successfully stewarded much of the research in the area and led to this survey. 

In Split Manufacturing, the key concept is to \textit{split} the circuit in two distinct parts before manufacturing, one containing the transistors and some routing wires, and the other containing the remaining routing wires. These parts are then fabricated in different foundries. The anatomy of an IC is illustrated in Figure~\ref{fig:ic_anatomy}, containing two set of layers, the bottom layer where the transistors are built, called Front end of the Line (FEOL), and the top layer where the metal layers are built for routing purposes, called Back end of Line (BEOL). The metal layers are referred as MX, where X is the level of the layer. M1 is the lowest layer at the bottom of the stack. Connections between metals are made by vias referred as VX, following the same naming scheme for metals. In Split Manufacturing, the FEOL is first manufactured in a high-end foundry, and later the BEOL is stacked on top of it by a second (and possibly low-end) foundry. This process requires electrical, mechanical, and/or optical alignment techniques to ensure the connections between them. Additionally, FEOL and BEOL technologies have to be compliant with each other \cite{Vaidyanathan2014} regarding the rules for metal/via dimensions where the split is to be performed. The split can be performed at the lowest metal layer (M1) or at higher layers, for which trade-offs are established between attained security and overheads. If the BEOL and FEOL technologies are vastly different from one another, Split Manufacturing may incur heavy overheads.

In this work, the focus is on the Split Manufacturing technique. As described above, Split Manufacturing can tackle threats that occur during the fabrication. It avoids overproduction, reverse engineering (to some extent) and unwanted modifications, limiting the capability of attackers. In Section II, we provide a background and more in-depth explanation of the technique. We address security threats in Section \ref{sec:bck_gnd}, demonstrating the potential vulnerabilities found in split circuits and describing the state-of-the-art attacks proposed until the present day. In Section \ref{sec:defenses}, the security of split circuits is discussed, showing how it can be improved using enhancement techniques. Future trends and lessons learnt are discussed in Section \ref{sec:trends}. Finally, our conclusions are presented in Section \ref{sec:conclusion}.

 \Figure[t!](topskip=0pt, botskip=0pt, midskip=0pt)[width=.9\linewidth]{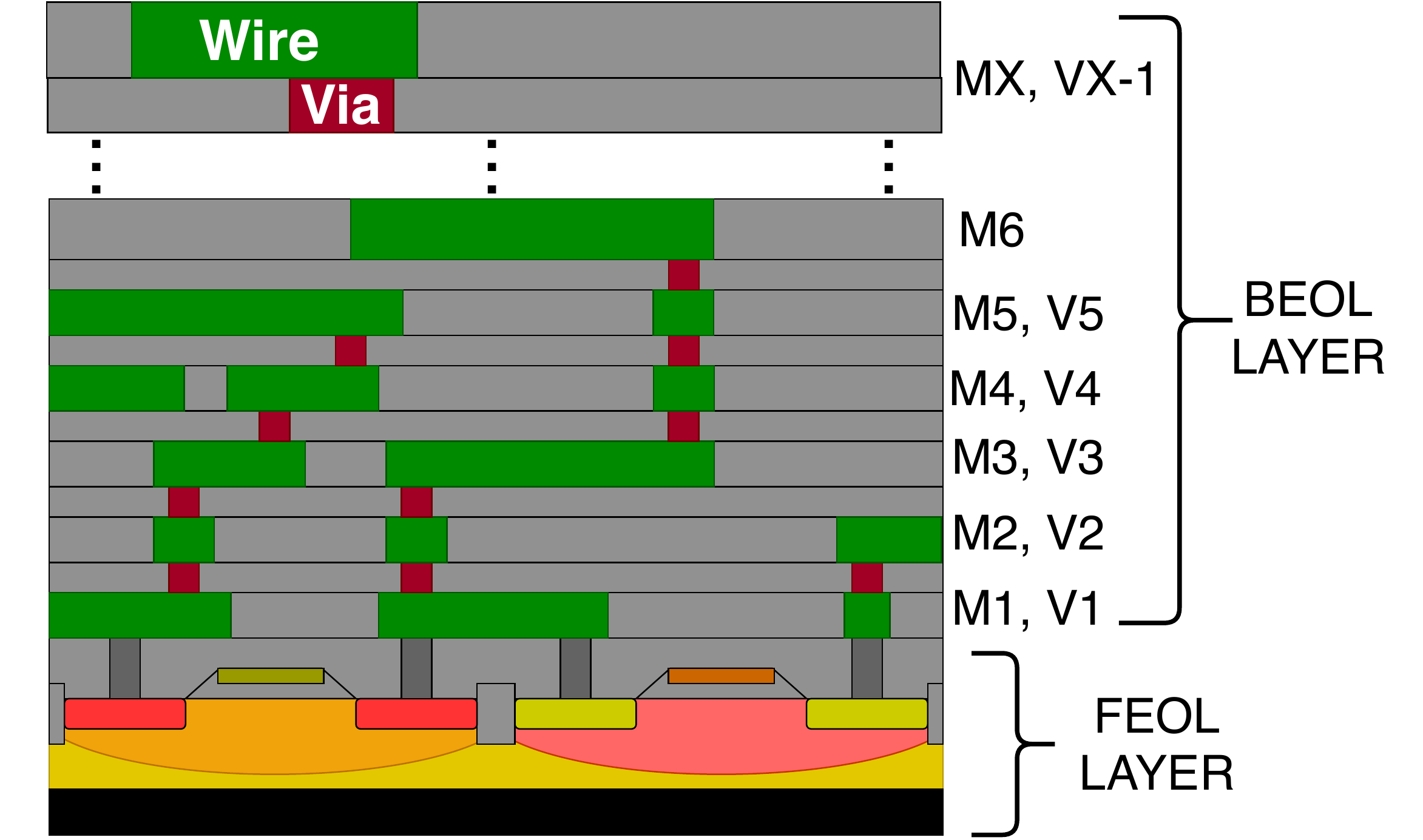}
{Anatomy of an integrated circuit (adapted from \cite{Rajendran2013}). \label{fig:ic_anatomy}}
 
 \section{Split Manufacturing: Background} \label{sec:bck_gnd}
 
As mentioned before, in order to have access to advanced technologies, many design companies have to outsource their IC manufacturing to untrusted high-end foundries. Protecting their designs against threats that may occur during manufacturing is a concern. Designs can be protected by applying the Split Manufacturing technique, thus combating all threats highlighted in blue in Figure~\ref{fig:cf_tax}.
  
Split Manufacturing protects a design by hiding sensitive data from the untrusted foundry. This is achieved by splitting the IC into two parts before manufacturing, a horizontal cut that breaks the circuit into one part containing the transistors and some (local) routing wires, and another containing only routing wires. These parts are termed FEOL and BEOL. 

As the FEOL and the BEOL of an IC are built sequentially, first FEOL and then BEOL, this characteristic enables the Split Manufacturing technique. Since the FEOL contains the transistors and possibly a few of the lowest ultra-thin metal layers -- the most complex parts of an CMOS process \cite{kikkawaJoshi} --, it is logical to seek to use a high-end foundry for its manufacturing, even if said foundry is not trusted. Completing the IC can then be done in a trusted low-end foundry, where the BEOL is stacked on top of the FEOL. Split Manufacturing was successfully demonstrated in \cite{Vaidyanathan2014, Vaidyanathan2014a, HillRKarmazin}, where designs were manufactured with \textasciitilde 0\% of faults, and are reported to present a performance overhead of about 5\%. Therefore, the technique is, in principle, feasible and available for design companies to make use of, such that they can utilize advanced foundries without fully exposing their layouts during manufacturing.
  
However, there are many caveats to Split Manufacturing. The technique can be successfully applied only if the technologies used to build the FEOL and BEOL are ``compatible''. In theory, a layout can be split at any layer if the chosen layer presents a good interface between FEOL and BEOL. However, since advanced technologies utilize the dual-damascene fabrication process, the layout can only be split on metal layers \cite{ddi}. Thus, the FEOL cannot terminate in a via layer. The dual-damascene process is characterized by patterning the vias and trenches in such a way that the metal deposition fills both at the same time, i.e., via-metal pairs (e.g., V1 and M2) must mandatorily be built by the same foundry. 
   
Two technologies are said to be compatible with each other if there is a way for a BEOL via to land on the FEOL uppermost layer while respecting all design rule checks (DRCs) of both technologies. DRCs are used to guarantee the manufacturability and functionality of an IC, and are defined with respected to the characteristics of the materials utilized and to tolerance ranges of the manufacturing processes (e.g., polishing, patterning, and deposition). These rules encompass minimum enclosure, width, spacing, and density checks. Modern technologies have several options for via shapes -- as long as one via shape is valid, the technologies are compatible for Split Manufacturing purposes. However, in practice, to keep the overhead of the technique under control, an array of via shapes must be feasible, thus providing the physical synthesis with a rich selection for both power and signal routing. 

According to \cite{Vaidyanathan2014}, compatibility between two technologies can be generalized by enclosure rules as in Eq.~\ref{eq:drc_1}, where MW.U.x is the minimum width of Mx on untrusted foundry, VW.T.x is the minimum width of Vx on trusted foundry and EN.T.x is the minimum enclosure on trusted foundry. These rules are illustrated in Figure \ref{fig:drc_mdim}, where the left side of the image portrays a cross section view and the right side shows the top view. According to Figure \ref{fig:drc_mdim}, the minimum enclosure width, Mx.EX.Vx, must be compatible between the two foundries. In modern technologies, Equation~\ref{eq:drc_1} is no longer sufficient as it does not capture the intricate rules for vias and line endings (enclosure from 1 side, 2 sides, 3 sides, T-shaped/hammerheads, etc.).

 \begin{equation} \label{eq:drc_1}
   MW.U.x \geq VW.T.x + (2 EN.T.x)
 \end{equation}

\Figure[t!](topskip=0pt, botskip=0pt, midskip=0pt)[width=0.95\linewidth]{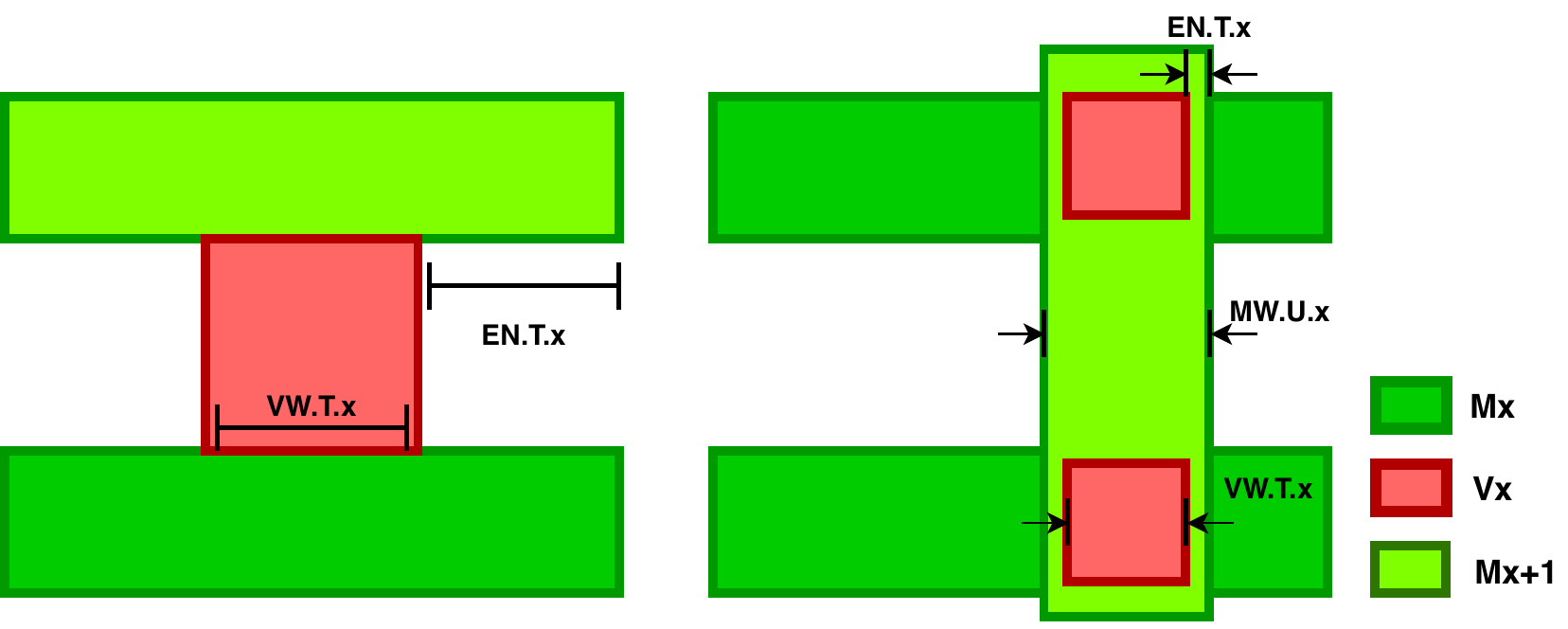}
{Compatibility rules between FEOL and BEOL (adapted from \cite{Vaidyanathan2014}). \label{fig:drc_mdim}}

Split Manufacturing also presents challenges on the design flow front, which is illustrated in Figure~\ref{fig:split_sm_flow}. An in-house team designs the circuit, from RTL to layout. Most likely, the layout contains IPs obtained from third parties. Depending on the metal layer where the layout is to be split, it may affect existing IP. Logic and memory IP may use higher metal layers -- memories typically require 4 to 5 metal layers, while standard cells typically require 2 metal layers --, limiting where the split can be done. Standard cells and memories have to be re-designed if they use metal layers that will be split, a grave challenge that may render Split Manufacturing significantly harder to execute.

Still referring to Figure~\ref{fig:split_sm_flow}, the FEOL and BEOL are generated using a hybrid process design kit (PDK), and then later split to be manufactured. After splitting the layout correctly, the FEOL is first manufactured in a high-end foundry, and later the BEOL is stacked on top of it by a second (and possibly low-end) foundry. 

\Figure[t!](topskip=0pt, botskip=0pt, midskip=0pt)[width=1\linewidth]{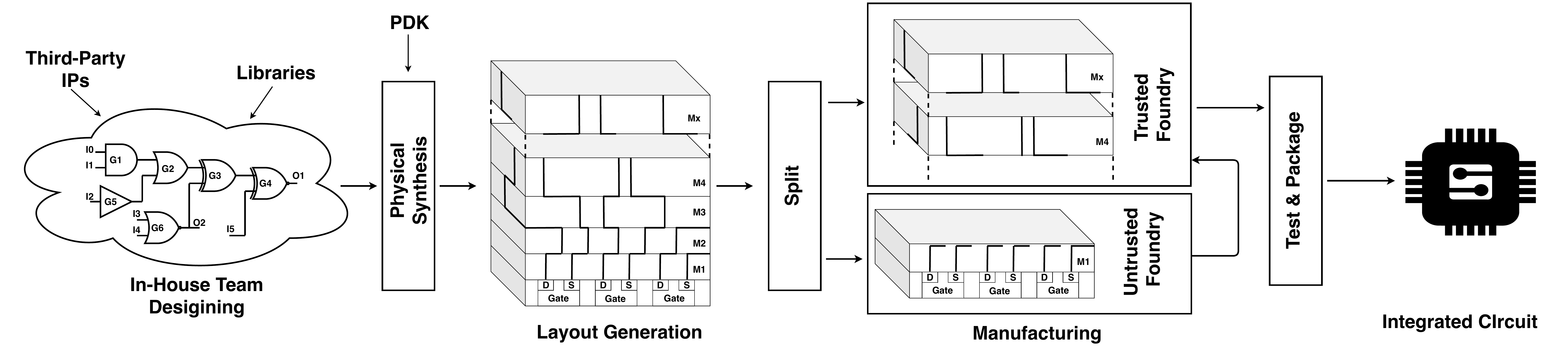}
{Split Manufacturing Design Flow. \label{fig:split_sm_flow}}

Even by splitting the layout, it is often argued that the FEOL exposes enough information to be exploited. Attacks towards the FEOL can effectively retrieve the BEOL connections by making educated guesses. The efficiency of the guessing process is inherently linked to the threat model assumed, which determines the information the attacker possesses to begin with. The literature describes two distinct threat models:
  
  \begin{itemize}
      \item \textbf{Threat model I:} an attacker located at the untrusted foundry holds the FEOL layout and wants to retrieve the BEOL connections.
      
      \item \textbf{Threat model II:} an attacker located at the untrusted foundry holds the entire gate-level netlist that is assumed to be provided by a malicious observer. The attacker here still holds the FEOL layout and wants to retrieve the BEOL connections. \cite{Imeson2013}.
  \end{itemize}
  
It is important to emphasize that the second threat model completely nullifies the security introduced by Split Manufacturing. Possessing the gate-level netlist makes reverse engineering the layout a trivial task, as if the attacker held the entire layout, not only the FEOL. Assuming the attacker has knowledge about the netlist challenges the integrity of the design company itself. If a rogue element exists inside the design company, other representations of the design could be equally stolen, such as the register-transfer level (RTL) code of the design, or even the entire layout (including the BEOL). It could be argued that this vulnerability is so severe that Split Manufacturing virtually stops making sense. For this reason, threat model I is the focus in this work. However, as our goal is to present a comprehensive survey, related works that use threat model II will be covered as well. 
  
Assuming threat model I, an attacker already knowing all the layers that make up the FEOL, is now interested in retrieving the BEOL connections to recreate the full design (or as close as possible). The commonly used assumption is that attackers are powerful and work within the untrusted foundry in some capacity. Thus, the attackers have deep understanding about the technology. Extracting the (still incomplete) gate-level netlist from a layout is, therefore, a trivial task for the attacker.
  
Many approaches to retrieve the BEOL connectivity have been proposed, several of which are termed \textit{proximity attacks} \cite{Rajendran2013,Wang2017,Yang2018}. Since EDA tools focus on optimizing power, performance, and area (PPA), the solution found by a placement algorithm (that uses heuristics internally) tends to place connected cells close to one another as this will, in turn, reduce area, wirelength, and delay. Therefore, finding the correct missing connections between FEOL and BEOL can be done by assessing input and output pins that are in proximity (thus, the name proximity attack). The more input and output pins to connect, the higher is the probability to make a wrong connectivity guess. Thus, a higher level of security is achieved by splitting the circuit at the lowest metal layer possible.
  
As a promising technique to enhance the security of ICs in this era of fabless chip design, Split Manufacturing still faces some enormous challenges:
  
  \begin{itemize}
      \item[] \textit{Logistical challenge:} Split Manufacturing is not presently incorporated into the IC supply chain. Finding foundries with compliant technologies that are willing to work with each other is not trivial.
      
      \item[] \textit{Technological challenge:} even within compliant technologies, non-negligible overheads can be introduced if they are vastly different\footnote{For a thorough discussion and silicon results on BEOL-related overheads, please refer to \cite{pagliarini}.}. In the worst-case scenario, it can make routing impossible. Thus, this fact narrows down the technology choices available and feasibility of certain layers as candidates for splitting.
      
      \item[] \textit{Security challenge:} the attained security of straightforward Split Manufacturing is still under debate. Attacks towards the FEOL can be effective, where the hidden connections can be retrieved.

  \end{itemize}
  
For the purpose of this survey, we categorized related works in the literature as attacks and defenses. In \textit{attacks}, authors proposed new attack models or modifications of existing attacks in order to improve their effectiveness. In \textit{defenses}, authors proposed new techniques to use together with Split Manufacturing in order to improve its security level.
 
\section{Attacks on Split Manufacturing} \label{sec:sec_thrts}

The Split Manufacturing technique was developed to protect ICs against threats related to manufacturing in potentially untrusted foundries. In practical terms, to split the layout means to hide some connections from the untrusted foundry. The security provided by Split Manufacturing is based on the fact that the attacker in the FEOL foundry cannot infer the missing BEOL connections. This assumption, however, was challenged by several works where authors proposed attack approaches that can potentially retrieve the missing connections with varying degrees of success. In the text that follows, we present works that proposed Split Manufacturing attacks. These attacks are discussed in chronological order as compiled in Table~\ref{tab:atcks_desc}. For each studied attack, we reported the related threat model, attack type, novelty, and benchmark circuits used in experiments. Additionally, we reported the largest and average size of the circuits utilized in each work (measured in number of gates). The circuit size is an important metric when analyzing the Split Manufacturing technique because its effectiveness is often proportional to circuit size. In Table~\ref{tab:atcks_effect}, we compiled results for when the smallest and largest studied circuits are under attack. Also, we reported the circuit size and, if defined, the split layer that the author selected to split the circuit at.

The first reported attack is by Jeyavijayan \etal~and is described in \cite{Rajendran2013}. In this work, the authors assumed that naive Split Manufacturing (i.e., splitting a layout without care for the connections) is inherently insecure. They introduced the concept of proximity attacks that exploits ``vulnerabilities'' introduced by EDA design tools. Since EDA tools focus on optimizing for PPA, the solution found by a placement algorithm tends to place logically connected cells close to one another so they become physically connected during routing. Therefore, the distance between output-input pairs can be used as a metric to recover the missing BEOL connections.

\Figure[t!](topskip=0pt, botskip=0pt, midskip=0pt)[width=0.90\linewidth]{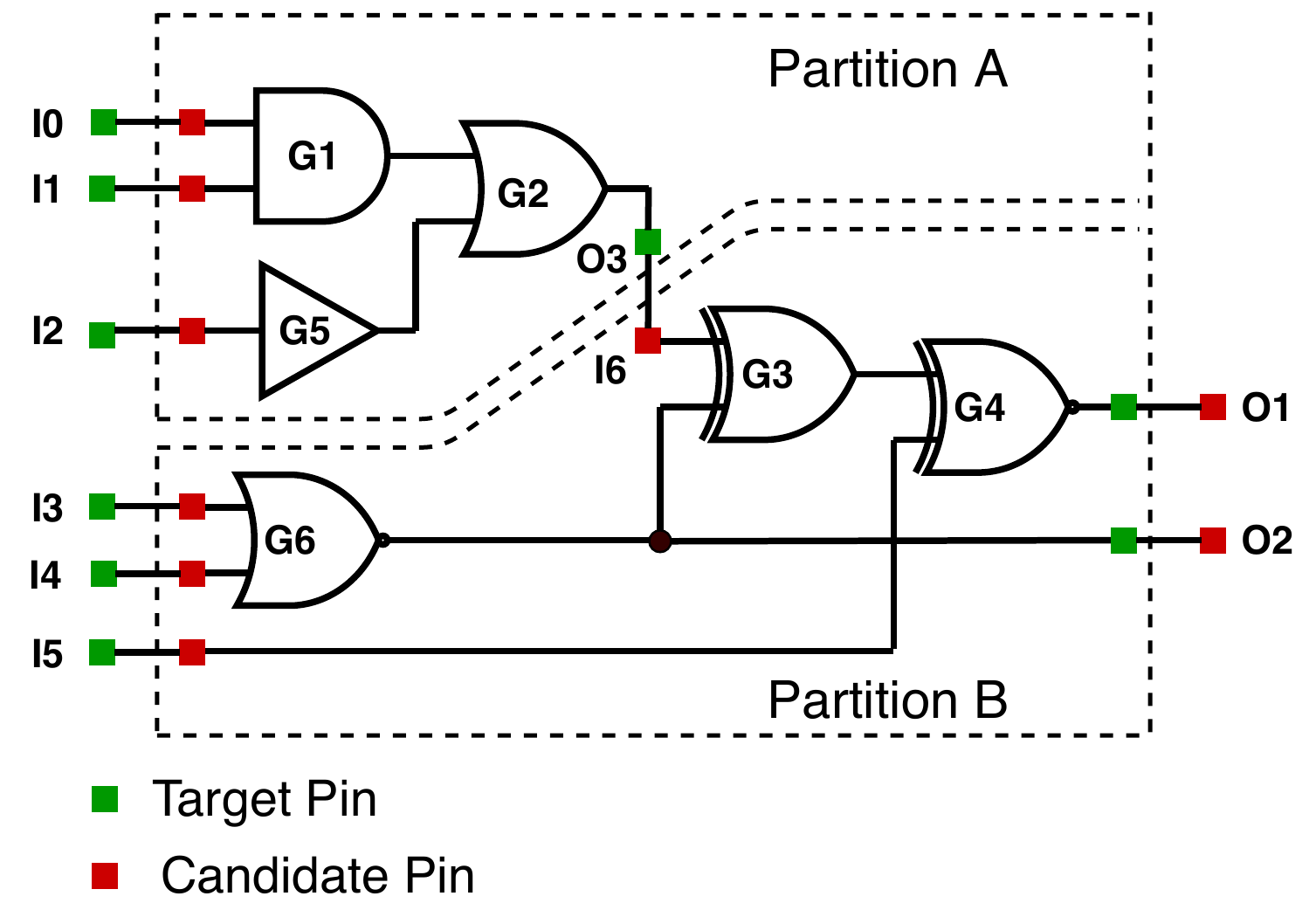}
{Example of a partitioned circuit. \label{fig:cir_part}}

Designs are commonly partitioned during physical implementation, i.e., separated into small logical blocks with few connections between them. That way, the designer has total control of the floorplaning regarding the placement of blocks. This approach also allows for blocks to be implemented separately and later integrated, creating a sense of parallelism in the design flow which can reduce the overall time required for executing physical synthesis. Consider as an example the circuit illustrated in Figure~\ref{fig:cir_part} which contains two partitions, A and B, each with 3 gates. Partition A has 3 input pins and one output pin, while partition B has three input pins and two output pins. The partitions have only one connection between them, connecting the output pin of gate G2 with one of the inputs of gate G3. Consider a target output pin from partition A $P_{x,A,out}$ and its corresponding candidate input pin from partition B $P_{x,B,in}$. During placement, the EDA tool will attempt to place the pin $P_{x,A,out}$ as close as possible to $P_{x,B,in}$, possibly closer than any other pin in partition B. Using this insight, an attacker may recover the missing connections in the FEOL layout, performing then a proximity attack. The authors argued that their proposed attack flow is successful due to being able to leverage the following ``hints'' provided by the EDA tools:
  
  \begin{itemize}
      \item[] \textit{Hint 1 - Input-Output Relationship}: partition input pins are connected either to another partition output pin or to an input port of the IC (i.e., input to input connections are excluded from the search space).
 
      \item[] \textit{Hint 2 - Unique Inputs per Partition}: input-output pins between partitions are connected by only one net. If a single partition output pin feeds more than one input pin, the fan-in and fan-out nodes are usually placed within the partitions (i.e., one-to-many connections are ruled out from the search space). 

      \item[] \textit{Hint 3 - Combinational Loops}: in general, only very specific structures are allowed to utilize combinational loops (e.g., ring oscillators). These structures are very easy to identify. In most cases, random logic does not contain combinational loops (i.e., connections that would lead to combinational loops can be excluded from the search space).
      
  \end{itemize}
  
An attacker can correctly connect a target pin to a candidate pin by identifying the closest pin from a list of possible candidates. The list of possible candidates is created by observing the hints mentioned above. A possible candidate pin is an unassigned output pin of another partition and an unassigned input port of the design. Then, a minimum distance metric is used to connect the pins based on the previously discussed heuristic behavior of EDA tools.
  
In Algorithm \ref{alg:prox_jevs}, we described the proximity attack detailed in \cite{Rajendran2013}. The input to the algorithm is the FEOL layout, from which the information about unassigned input-output ports can be derived. The algorithm does not describe the specifics of how to derive a netlist from a layout. However, the complexity of this task is rather straightforward. It is assumed that the attacker possesses information about both the PDK and the standard cell library. In many cases, the untrusted foundry is the actual provider of both\footnote{For the PDK, it is very natural that it is created by the untrusted foundry itself. For standard cell libraries, the cells might be designed by the foundry or by a third-party licensed by the foundry. In either case, the effort to revert a layout to a netlist remains trivial.}. From there, a layout in GDSII or OASIS format can be easily reverted to a netlist by any custom design EDA tool. 

\begin{algorithm}[h]
\DontPrintSemicolon
  
  \KwInput{FEOL layers}
  \KwOutput{Netlist with BEOL connections}
  Reverse engineer FEOL layers and obtain partitions;
  
   \While{Unassigned partition pins or ports exist}
   {
   Select arbitrary unassigned pin/port as a targetPin;
   
   	ListOfCandPins = BuildCandPinsList(targetPin);
   	
   	Select candPin from ListOfCandPins that is closest to targetPin;
   	
   	Connect targetPin and candPin;
   	
   	Update netlist;
   }
   
   \textbf{Return:} netlist

   \textbf{BuildCandPinsList}(targetPin)
   
   \KwInput{targetPin $P_{X,i,in}$}
  \KwOutput{CandPins for targetPin}
  
  CandPins = Unassigned output pins of other partitions + unassigned input ports of the design;
  
  \For{\textbf{each} $Pin_J \in$ CandPins}
   {
    \If{CombinationalLoop(targetPin, $Pin_J$)}
    {
        CandPins -= $Pin_J$;
    }
   }
   \textbf{Return:} CandPins
\caption{Proximity attack}
\label{alg:prox_jevs}
\end{algorithm}

From the gate-level netlist, the algorithm chooses an arbitrary \textit{TargetPin} from the unassigned partition input pins and output ports, creates a list of possible \textit{CandidatePins}, and then connects the \textit{TargetPin} to the closest pin in this list. After each connection, the netlist is updated. This procedure is repeated until all unassigned ports are connected. When the procedure is over, the attacker obtains the possible missing BEOL connections. If all guesses were correct, the original design has been recovered and Split Manufacturing has been defeated.

 \begin{table*}[htb]
 \rowcolors{2}{gray!25}{white}
    \centering
    \caption{Threat Models, Attacks, and Metrics.}
    \begin{tabular}{ccclp{4.0cm}lp{1.7cm}p{1.2cm}}
    \hline \hline \\ [-1.5ex]
        \textbf{Work} & \textbf{Year} & \parbox{1cm}{\textbf{Threat} \\\textbf{model}} & \textbf{Attack type } & \textbf{Novelty} & \textbf{Benchmark suite(s)} & \parbox{2cm}{\textbf{Largest circuit} \\\textbf{size (gates)}} & \parbox{2cm}{\textbf{Avg. circuit} \\\textbf{size (gates)} } \vspace{1pt} \\ 
        \hline
         \cite{Rajendran2013} & 2013 &  I & Proximity & Attack Based on Proximity & ISCAS\textquotesingle85 & 3.51k & 1288\\
         \cite{Magana2016} & 2016 &  II & Proximity  & \parbox{4.5cm}{Placement and routing proximity \\ used in conjunction} & ISPD\textquotesingle11 & 1.29M & 951k \\
         \cite{Wang2018} & 2018 &  I & Proximity & \parbox{4.5cm}{Network-Flow-Based with  Design \\ Based Hints} & ISCAS\textquotesingle85 \& ITC\textquotesingle99 & 190.21k & 9856\\
         \cite{Zhang2018} & 2018 & I & Proximity & \parbox{4.5cm}{Proximity Attack Based on Machine \\ Learning} & ISPD\textquotesingle11 &  1.29M & 951k\\
         \cite{Chen2019b} & 2019 &  I & SAT  & SAT Attack without Proximity Information & ISCAS\textquotesingle85 \& ITC\textquotesingle99 & 190.21k & 9856\\
         \cite{Chen2019a} & 2019 & I & SAT & \parbox{4.5cm}{SAT attack dynamically adjusted \\based on proximity information} &  ISCAS\textquotesingle85 \& ITC\textquotesingle99 & 190.21K & 9856\\
         \hline \hline
    \end{tabular}
    
    \label{tab:atcks_desc}
\end{table*}
  
Algorithm \ref{alg:prox_jevs} was originally applied to the ISCAS\textquotesingle85 \cite{ISCAS85} suite of benchmark circuits. These circuits were originally selected and published to help in comparing automatic test pattern generation (ATPG) tools. Due to the small size of these circuits, they may not be the best option to assess the effectiveness of Split Manufacturing. The difficulty to retrieve the BEOL connections is directly proportional to the size of the circuit. The authors reported an average effectiveness of 96\% of Correct Connection Rate (CCR) across all the benchmarks considered. For the c17 circuit (smallest circuit in the ISCAS\textquotesingle85 suite with only 6 gates), all the connections were retrieved correctly, thus demonstrating that the algorithm is capable of retrieving the missing BEOL connections. In Table~\ref{tab:atcks_effect}, we highlight the best and worst results in terms of CCR.
  
Jeyavijayan \etal \cite{Rajendran2013} were the first to question the security of straightforward Split Manufacturing. Their proximity attack showed promising results, even if the considered benchmark circuits were rather small in size. This was the starting point for other studies proposing different attacks to Split Manufacturing in an attempt to retrieve the missing BEOL connections. Improvements over the original proximity attack, as well as other attacks, are compiled in Table~\ref{tab:atcks_desc}. 

The effectiveness of the proximity attack utilizing distance of unassigned pins alone as metric to find missing BEOL connections was questioned by Maga\~na \etal \cite{Magana2016}. The authors proposed to utilize both placement and routing information in augmented proximity attacks. For their results, large-sized circuits from the ISPD-2011 routability-driven placement contest \cite{ispd_bench} were used. These benchmarks are better representatives of modern circuits as they contain 9 metal layers and up to two million nets in a design. Thus, in an attempt to increase the success rate of the attack for large-sized circuits, they proposed routing-based proximity in conjunction with placement-centric proximity attacks.
 
 \begin{table*}[htb]
 \rowcolors{2}{gray!25}{white}
    \centering
       \caption{Benchmark Size and Comparison of Attack Results.}
    \begin{tabular}{cllcllc}
    \hline \hline \\ [-1.5ex]
        \textbf{Work} & \textbf{Benchmark} & \textbf{Attack} & \textbf{Split Layer} & \textbf{Size (In Gate Count)} & \textbf{Metric} & \textbf{Result}  \vspace{5pt} \\
        \hline
        \cite{Rajendran2013} & c17 & Proximity & Not Defined & 6 & CCR(\%) & 100 \\
         \cite{Rajendran2013} & c7552 & Proximity & Not Defined & 3513 & CCR(\%) & 94  \\
         \cite{Magana2016} & Superblue 1 & Placement Proximity & M2 & 847k & \% Match in List & 12.84 \\
         \cite{Magana2016} & Superblue 1 & Placement Proximity & M2 & 847k & CCR(\%)  & 5.479 \\
         \cite{Magana2016} & Superblue 1 & Routing Proximity & M2 & 847k & \% Match in List & 71.08 \\
         \cite{Magana2016} & Superblue 1 & Routing Proximity & M2 & 847k & CCR(\%)  & 0.651 \\
         \cite{Magana2016} & Superblue 1 & Overlap (P\&R) Proximity & M2 & 847k & \% Match in List & 13.05 \\
         \cite{Magana2016} & Superblue 1 & Overlap (P\&R) Proximity & M2 & 847k & CCR(\%)  & 3.977 \\
         \cite{Magana2016} & Superblue 1 & Crouting Proximity & M2 & 847k & \% Match in List & 82.08 \\
         \cite{Magana2016} & Superblue 1 & Crouting Proximity & M2 & 847k & CCR(\%)  & 0.651 \\
         \cite{Wang2018}  &  c7552 & Network-flow Based Proximity & Not Defined & 3513 & CCR(\%) & 93 \\
         \cite{Wang2018}  &  c7552 & Proximity & Not Defined & 3513 & CCR(\%) & 42 \\
         \cite{Wang2018}  &  B18 & Network-flow Based Proximity & Not Defined & 94249 & CCR(\%) & 17 \\
         \cite{Wang2018}  &  B18 & Proximity & Not Defined & 94249 & CCR(\%) & < 1 \\
         \cite{Zhang2018} & Superblue 1 & Proximity & M6 & 847k & \% Match in list & 33.40 \\
         \cite{Zhang2018} & Superblue 1 & Proximity & M6 & 847k & CCR(\%) & 0.76 \\
         \cite{Zhang2018} & Superblue 1 & ML & M6 & 847k & \% Match in list & 83.12 \\
         \cite{Zhang2018} & Superblue 1 & ML & M6 & 847k & CCR(\%) & 1.91 \\
         \cite{Zhang2018} & Superblue 1 & ML-imp & M6 & 847k & \% Match in list & 74.65 \\
         \cite{Zhang2018} & Superblue 1 & ML-imp & M6 & 847k & CCR(\%) & 2.11 \\
         \cite{Zhang2018} & Superblue 1 & ML-imp & M4 & 847k & \% Match in list & 75.45 \\
         \cite{Zhang2018} & Superblue 1 & ML-imp & M4 & 847k & CCR(\%) & 2.58 \\
         \cite{Chen2019b} & c7552 & SAT Attack & Not Defined & 3513 & Logical Equivalence(\%) & 100 \\
         \cite{Chen2019b} & B18   & SAT Attack & Not Defined & 94249 & Logical Equivalence(\%) & 100 \\
         \cite{Chen2019a} & c7552 & Improved SAT Attack & Not Defined & 3513 & Logical Equivalence(\%) & 100 \\
         \cite{Chen2019a} & B18   & Improved SAT Attack & Not Defined & 94249 & Logical Equivalence(\%) & 100 \\
    \hline \hline
    \end{tabular}
    \label{tab:atcks_effect}
\end{table*}
 
A key difference present in \cite{Magana2016} is that this work utilizes a different threat model (model II), claiming that the untrusted foundry possesses information about the entire place \& routed netlist (as well as the FEOL layout). This assumption is hard to reason if the attacker's intent was to overproduce the IC or pirate the IP. For these goals, clearly, this assumption is unnecessary. The attacker himself can, if he indeed possesses the netlist, perform his own physical synthesis and generate his own layout. The interest in reverse engineering the BEOL connections of the original design diminishes. Nevertheless, we report on the strategies employed by the authors of \cite{Magana2016} since they build on the approach proposed by \cite{Rajendran2013}.

\Figure[t!](topskip=0pt, botskip=0pt, midskip=0pt)[width=0.90\linewidth]{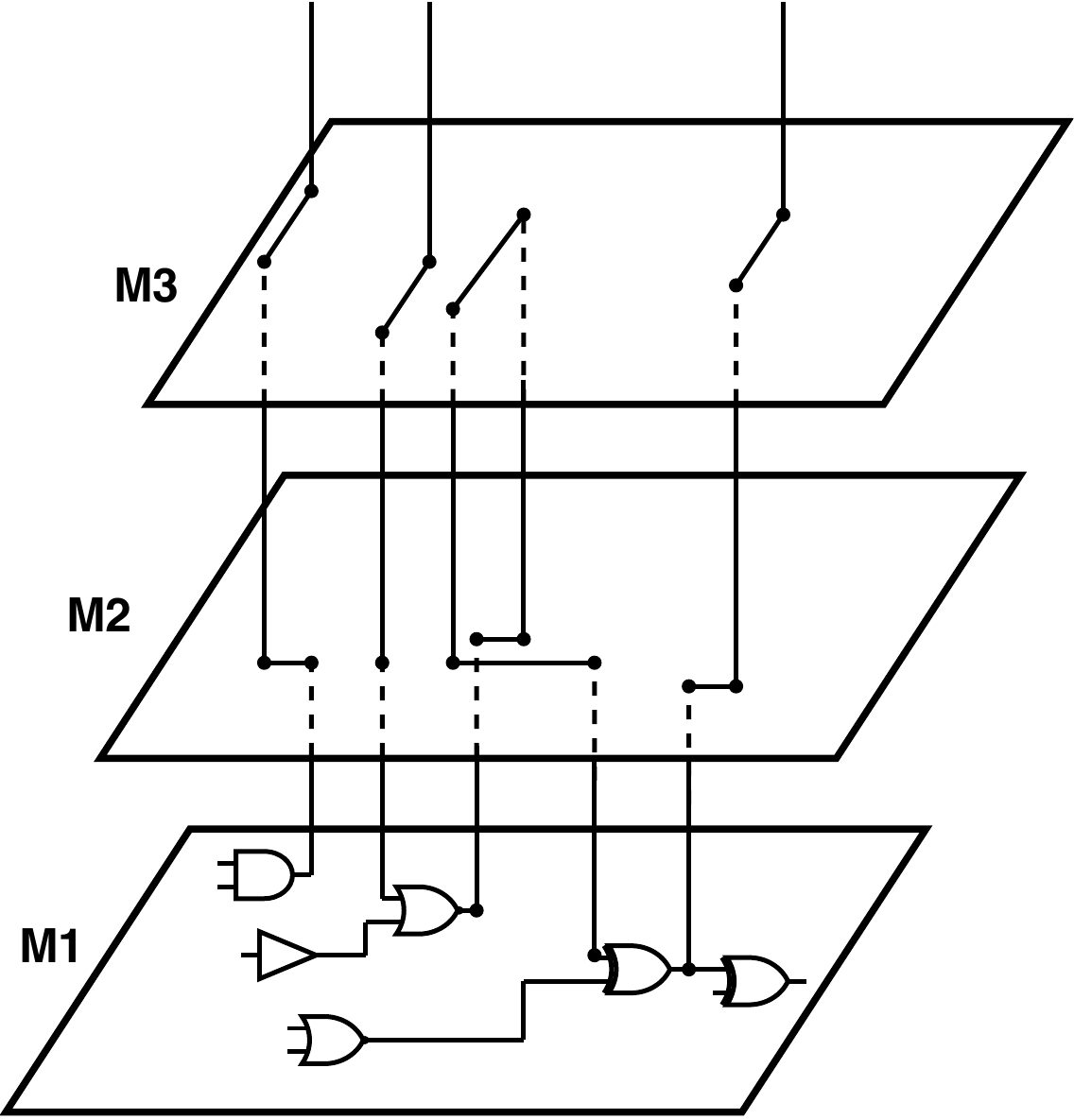}
{Representation of the routing for the first 3 metal layers of a simple circuit (adapted from \cite{Magana2016}). \label{fig:circ_abs}}

Regarding the attacks, the authors of \cite{Magana2016} proposed four different techniques to identify a small search neighborhood for each pin. The goal is to create a neighborhood that is small enough to make further pruning feasible, and therefore increase the likelihood of including the matching pins. The techniques are called \textit{placement proximity}, \textit{routing proximity}, \textit{crouting proximity} and \textit{overlap of placement and routing proximity}, and are described in the text that follows. The circuit illustrated in Figure~\ref{fig:circ_abs} is the example (before the split) that will guide the discussion on these four techniques.
 
\textit{Placement proximity} exploits the placement information of cells. Each split wire is taken from the pin location of the corresponding standard cell that is connected to it. A search neighborhood is defined as a square region centered around the corresponding pin with an area equal to the average areas of the bounding boxes (BB) in a typical design. The authors argued that it can also be measured based on BBs of the non-split wires in the design under attack, under the assumption that the number of wires that remain in the FEOL is also very large in practice. Let us consider the circuit illustrated in Figure~\ref{fig:circ_abs} as an example. If the split is done at M2, the search area defined using the placement proximity would contain three gates as illustrated in Figure~\ref{fig:prox_sa} (a). Thus, using the placement proximity search area, the most likely connections are illustrated by the green squares (candidate pins) and by the red squares (target pins). Note that the layer at which the layout is split does not affect the search area defined by placement proximity.
 
\textit{Routing proximity} exploits the routing information. First, for each split wire, pins are identified as the point where the wire is actually cut at the split layer, i.e., the via location. Next, a square area centered around those pins is defined. The size of the square area is defined based on the average BBs of the pins on that layer in the design. This procedure for identification results in different neighborhood sizes according to the split layer location, i.e., the search radius adapts to the routing resources of each layer. A search area defined using routing proximity is illustrated in Figure~\ref{fig:prox_sa} (b), highlighted in gray and containing a set of routing wires and its respective pins.  
 
\textit{Crouting proximity} takes into account routing congestion by exploiting the union of placement and routing proximity. The search area for each pin is defined in such way that the ratio of number of pins to the search area is equal across all the pins in the split layer. Thus, if a pin is located at a high routing congestion area, the search area will be expanded until the pin density in the new search area reaches a target value or the search area grows to four times its starting value. The starting value is set according to the split layer, set as the average of numbers of pins which fall within a BB. A search area defined using crouting proximity is illustrated in Figure~\ref{fig:prox_sa} (c). 
 
The last strategy proposed by \cite{Magana2016} also combines placement and routing information. It is referred to as \textit{Overlap of placement and routing proximities}. The concept here is to include a subset of pins identified by the placement proximity list which have their corresponding pins included in the routing proximity list. According to the author, intuitively, the overlap then identifies a subset of pins which may be more likely to point towards the direction of the matching pin. A search area defined using the overlap of placement and routing proximities is illustrated in Figure~\ref{fig:prox_sa} (d). 
 
 \Figure[t!](topskip=0pt, botskip=0pt, midskip=0pt)[width=1\linewidth]{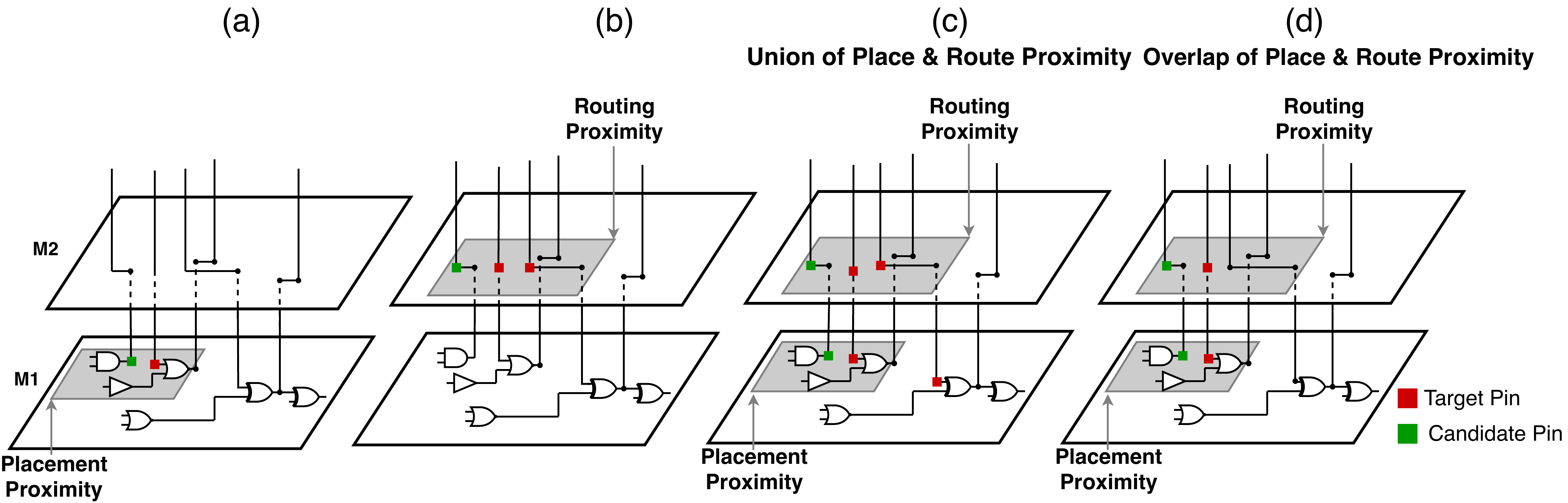}
{Multiple strategies for pin/connectivity search areas according to \cite{Magana2016}. \label{fig:prox_sa}}

Maga\~na et al. \cite{Magana2016} assessed each strategy using the benchmark circuit \textit{superblue1}. Different split layers were also considered. In Table~\ref{tab:atcks_effect}, we compiled the results for split layer M2. By comparing the results, it becomes clear that no strategy was able to recover 100\% of the missing BEOL connections. The best result was only 5.479\% of CCR. This is in heavy contrast with the findings of \cite{Rajendran2013}. However, as we previously noted, the circuit sizes differ by orders of magnitude.

According to the authors of \cite{Magana2016}, proximity alone is in no way sufficient to reverse engineer the FEOL. However, proximity attacks have merit as they can be used to narrow down the list of candidates to a significantly smaller size. Using crouting proximity, in 82.08\% of the cases, the search area defined contained the matched pin in the list of candidates. The authors also present results for a circuit split at M8. We opt not to show these results in Table~\ref{tab:atcks_effect}. Using the circuit \textit{superblue1} as an example, the number of unassigned pins when the circuit is split at M8 is only 1.2\% of the pins when split at M2. Therefore, the small number of unassigned pins to be connected overshadows the large circuit used for their experiments. It must also be emphasized that splitting a circuit in such higher layers is rather impractical since M8 tends to be a very thick metal reserved for power distribution in typical 10-metal stacks. There is very little value in hiding a power distribution network from an adversary that wants to pirate an IP. Once again, we opt not to show this result in our comparisons.
 
A network-flow based attack model towards flattened designs was proposed by Wang et. al \cite{Wang2018}. The authors argued that the proximity attack originally proposed by \cite{Rajendran2013} utilizes hints that can be used only by hierarchical designs, and that modern designs are often flattened \footnote{We highlight that best practices in circuit design have changed over the years. Hierarchical design was heavily utilized for many years, but it lost favor due to the difficulty in performing reasonable timing budgeting between the many blocks of a system. Thus, flattened designs are often used to facilitate timing closure.}. Based on the original proximity attack, they proposed a proximity attack utilizing five hints: physical proximity, acyclic combinational logic circuit, load capacitance constraint, directionality of dangling wires, and timing constraint. Note that the first two hints are already described by \cite{Rajendran2013} and \cite{Magana2016}. The three novel hints are described below:
 
 \begin{itemize}
     \item[] \textit{Load Capacitance Constraint}: gates can drive a limited load to honor slew constraints. Typically, maximum load capacitance is constrained and has a maximum value defined by the PDK and/or the standard cell characterization boundaries. Hence, an attacker will consider only connections that will not violate load capacitance constraints.
     
     \item[] \textit{Directionality of Dangling Wires}: routing engines tend to route wires from a source to a sink node along the direction of the sink node. Therefore, the directionality of remaining dangling wires at lower metal layers may indicate the direction of their destination cell with a high degree of certainty\footnote{Metals usually have preferred directions that alternate along the stack (i.e., if M1 is vertical, then M2 is horizontal). Therefore, this hint becomes more effective if the attacker can observe more than one routing layer of the FEOL}. An attacker can disregard connections in the other directions.
     
     \item[] \textit{Timing Constraint}: connections that create timing paths that violate timing constraints can be excluded. An attacker, through an educated guess of the clock period, can determine a conservative timing constraint and exclude any connections that would lead to slower paths. 
    
 \end{itemize}
 
The network-flow based attacked framework proposed by Yang \etal~considers two hints proposed by \cite{Rajendran2013} plus the aforementioned hints to create a directed graph $G = (V,E)$, where $V$ is a set of vertices and $E$ is a set of edges. The set ($V$) is composed by the set of vertices corresponding to the output pins $V_o$, and a set corresponding to the input pins ($V_i$), the source vertex ($S$), and the target vertex ($T$). The set $E$ consists of $E_{So}$, edges from $S$ to every output pin vertex, $E_{oi}$, edges from output pin vertices to input pin vertices, and $E_{iT}$, which includes edges from every input vertex to the target vertex. An example of this kind of representation is shown in Figure~\ref{fig:network_flow}, where (a) is the circuit with missing connections and (b) is the network-flow representation. The detailed problem formulation is omitted from this work. To find the connections, a min-cost network-flow problem is solved, where the decision variables are the flow $x_{i,j}$ going through each edge $(i,j) \in E$ . The authors utilized the Edmons-karp algorithm \cite{networkFlowAlg} to solve this problem. Complexity of the algorithm alone is given by $O(VE^2)$, however, in the worst case it is required to run the algorithm $V$ times; thus, the run-time of the complete network attack is given by $O(V^2E^2)$ in the worst case.
 
\Figure[t!](topskip=0pt, botskip=0pt, midskip=0pt)[width=0.90\linewidth]{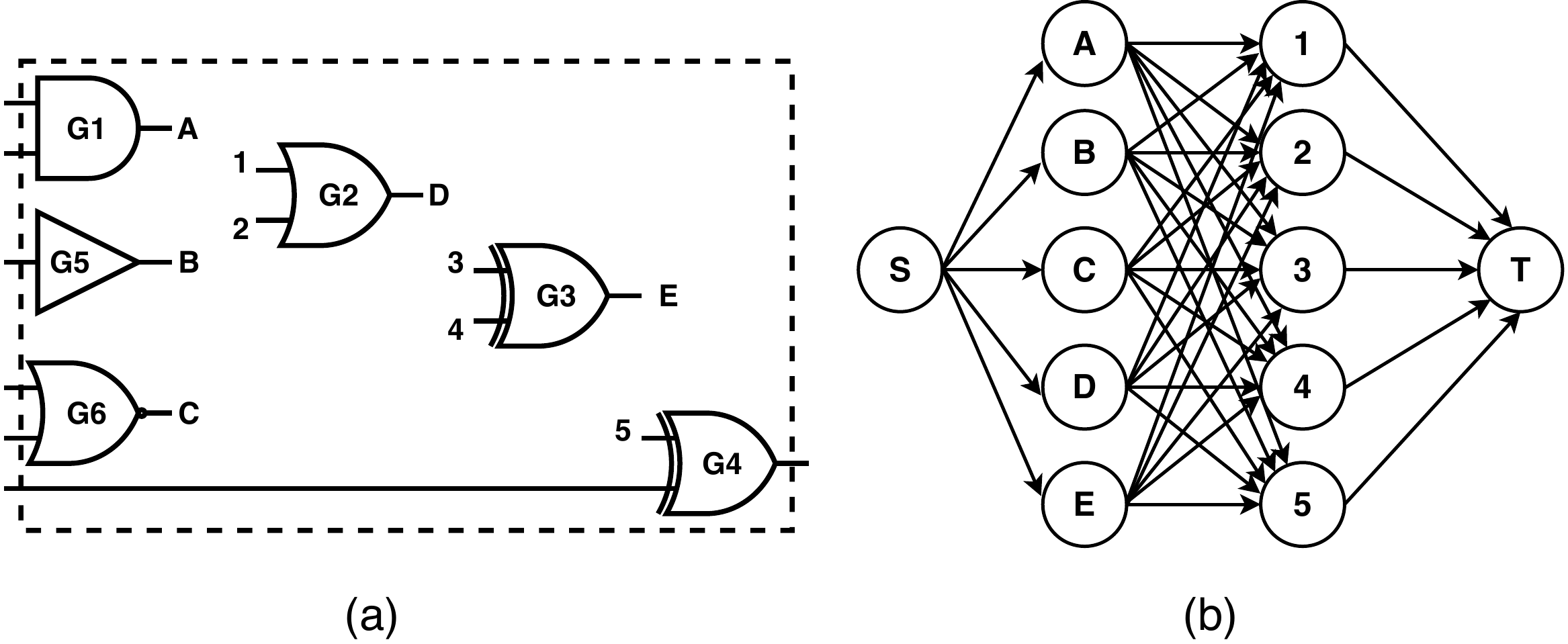}
{(a) Circuit with missing connections. (b) Network-flow model for inferring the missing connections. (Adapted from \cite{Wang2018}). \label{fig:network_flow}}
 
The network-flow approach was applied to ISCAS\textquotesingle85 and ITC\textquotesingle99 \cite{ITC99} benchmark circuits. The ITC\textquotesingle99 circuits were proposed as an evolution of the ISCAS\textquotesingle85 set, since the Test community acknowledged that newer and larger circuits were already in demand. For comparison, the authors applied both the original proximity attack and the network-flow attack to flattened designs. As shown in Table~\ref{tab:atcks_effect}, their network-flow proximity attack outperformed the original attack in terms of CCR. However, despite the evident improvement, the attack could only retrieve 17\% of the missing BEOL connections for a medium sized circuit (b18 from the ITC\textquotesingle99 suite).
 
A Machine Learning (ML) framework was used by Zhang \etal \cite{Zhang2018} in an attempt to improve the attack proposed in \cite{Magana2016}. The same setup as previously discussed was utilized. However, more layout features were incorporated in their ML formulation, including placement, routing, cell sizes, and cell pin types. 
 
A high-level overview of their modeling framework is shown in Figure~\ref{fig:ml_framework} (a). First, they create a challenge instance from the entire layout and only FEOL view. Next, for each virtual pin (point where a net is broken on the split layer), layout information is collected, including placement, routing, cell areas, and cell pin as illustrated in Figure~\ref{fig:ml_framework} (b). Using this information, samples are generated which are fed into the ML training process. Each sample carries information of a pair of virtual pins which may or may not be matched. Classifiers then are built by the ML framework using training samples. After training and building the regression model, cross validation is used for evaluation which ensures validation of the model is done on data samples which were not used for training. Their framework faces scaling issues when applied to lower split metal layers. An improved ML framework is then proposed as well, denoted by ML-imp, to solve the scaling issues.
 
For their experiments, Zhang \etal \cite{Zhang2018} utilize the ISPD\textquotesingle11 benchmark suite. They compare results from their previous work \cite{Magana2016} with their ML and ML-imp frameworks. However, they do not show results for lower split metal layers (e.g., M2). Instead, results are provided for M8, M6, and M4 splits. As pointed out before, utilizing higher layers for the split effectively shrinks the otherwise large circuits used in their experiments. A drastic reduction of unassigned pins is expected for such higher layers, as higher metal layers are used often for power routing, not for signal routing. Results for the \textit{superblue1} circuit are shown in Table~\ref{tab:atcks_effect}. Regarding recovering missing BEOL connections, ML and ML-imp could only retrieve around 2\%, therefore not showing a huge improvement over their previous work. However, search list area accuracy showed significantly better results when compared to their prior work. A caveat worth mentioning is that the proposed machine learning framework needs the entire layout during its modeling phase. This characteristic may, in an extreme case, nullify the applicability for an attacker that only holds the FEOL layout and cannot produce training samples from other sources.
 
 \Figure[t!](topskip=0pt, botskip=0pt, midskip=0pt)[width=1\linewidth]{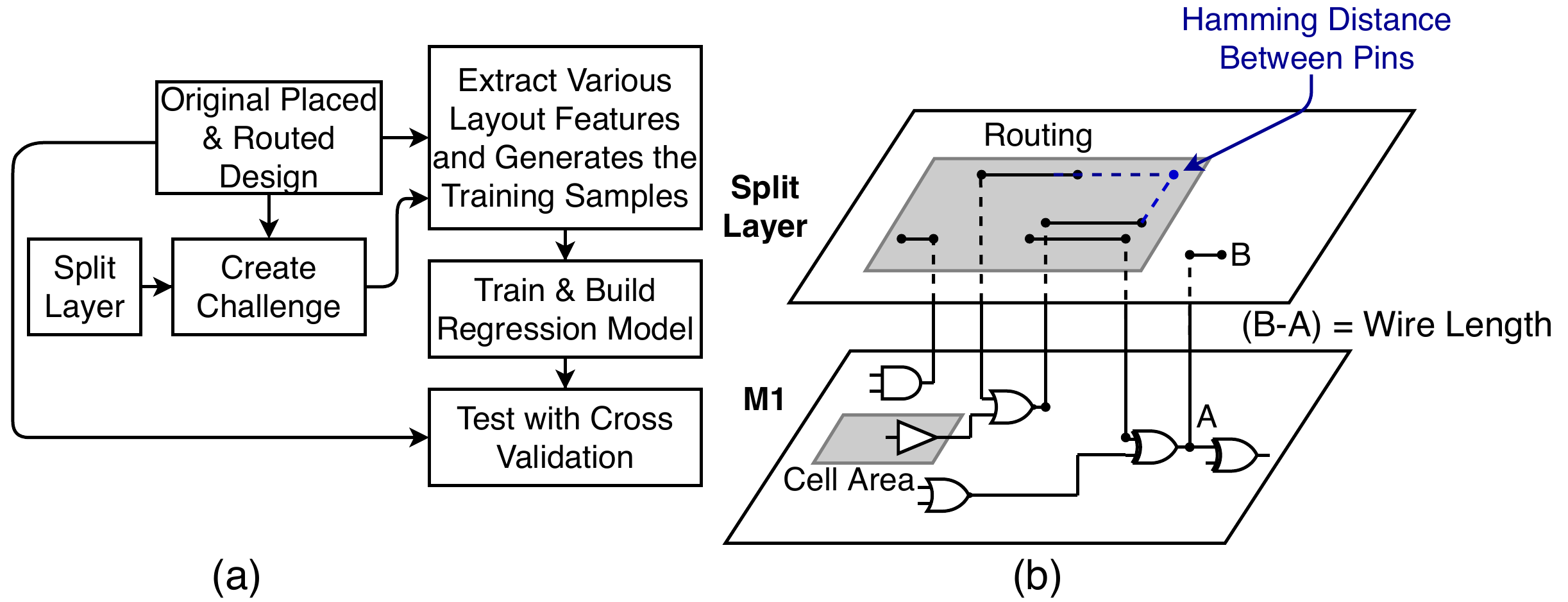}
{(a) Machine Learning Modeling by Zhang et al. \cite{Zhang2018}. (b) Few Exmples of Layout Features. \label{fig:ml_framework}}
 
Attacks using proximity information as a metric are not the only solution to recover missing BEOL connections. An effective methodology to apply a Boolean satisfiability based strategy is proposed by Chen \etal \cite{Chen2019b}. The authors claim that their attack methodology does not need (or depend on) any proximity information, or even any other insights into the nature of EDA tools utilized in the design process. The key insight in their work is to model the interconnect network as key-controlled multiplexers (MUX). Initially, all combinations of signal connections between the FEOL partitions are allowed, as illustrated in Figure~\ref{fig:mux_network}. First, a MUX network is created in order to connect all missing paths in the circuit. This MUX network leads to potential cyclic paths, thus, there is a possibility to generate many combinational loops during the attack process corresponding to incorrect key guesses. Therefore, constraints on the key values are generated in order to avoid activating the cyclic paths. The attack can be summarized in 4 steps: \textit{identification of all cyclic paths}, \textit{generation of cycle constraints}, \textit{cycle constraints optimizations}, and finally, \textit{SAT attack}. The authors utilize a SAT solver-based attack method derived from CycSat \cite{zhou_sat}. The SAT attack algorithm has as input the FEOL circuit with MUX network and a packaged IC that serves as an oracle. The algorithm outputs keys to be used in the MUXes such that correct BEOL connections are made. 

In reality, \cite{Chen2019b} presents a different interpretation of Threat model I since the attacker is assumed to possess a functional IC. This IC would then have to be available in the open market for the attacker to be able to purchase it. This characteristic severely narrows down the applicability of this SAT attack. For instance, ICs designed for space or military use will not be freely available, thus an oracle may not be known to the attacker.
  
Experimental results presented by \cite{Chen2019b} utilize ISCAS\textquotesingle85 and ITC\textquotesingle99 benchmark circuits. It has been shown that their attack could recover a logically correct netlist for all the studied circuits. However, there is a small clarification to be made that relates to what is a logically correct circuit. In Table~\ref{tab:atcks_effect}, two of those results are shown. For seven of the studied benchmarks (c1908, c2670, c5315, c7552, b14, b15, b17), the connections recovered are identical to the BEOL connections. For the remaining benchmarks, the recovered connections are not identical but logically equivalent to the original circuit. In practice, the logically equivalent circuit may present performance deviations from the original design. Matching the performance of the original design can be done by re-executing place and route using the logically equivalent gate-level netlist. Depending on the attack goal, it is possible that the attacker had already planned to re-execute the physical synthesis flow again (say, to resell the IP in a different form or shape). An attack that guarantees 100\% of logic equivalence of the recovered netlist is powerful enough, allowing attackers to copy and modify split layouts.
  
In order to increase the efficiency and capacity of the SAT attack proposed in \cite{Chen2019b}, the authors proposed two improvements in \cite{Chen2019a}. First, the size of the key-controlled interconnect network that models the possible BEOL connections is reduced. Second, after the MUX network is inserted into the FEOL circuit, the number of combinational cycles it induces in the design for incorrect key guesses should also be reduced. Proximity information is then exploited to achieve the proposed improvements. The improved SAT attack method which exploits proximity information showed significant reduction in the attack time and increase in the capacity. Same as in \cite{Chen2019b}, the circuits tested were 100\% recovered, as shown in Table~\ref{tab:atcks_effect}. 

\Figure[t!](topskip=0pt, botskip=0pt, midskip=0pt)[width=0.95\linewidth]{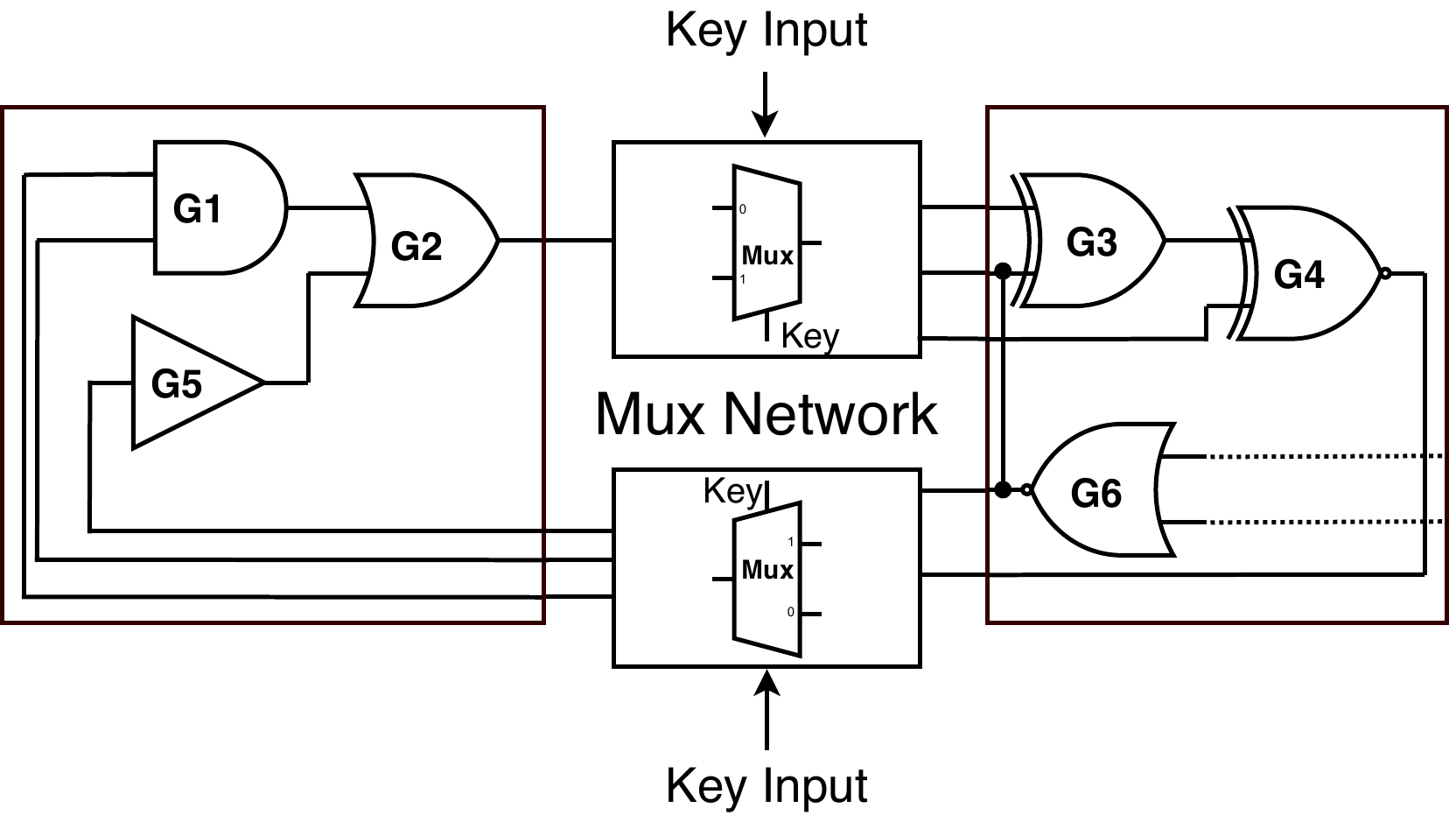}
{MUX network for a bipartitioned FEOL circuit (Adapted from \cite{Chen2019b}). \label{fig:mux_network}}

\begin{table*}[htb]
\rowcolors{2}{gray!25}{white}
    \centering
    \caption{Split Manufacturing Defenses.}
    \begin{tabular}{cp{1cm}ccp{2.2cm}p{2cm}l}
    \hline \hline \\ [-1.5ex]
      \textbf{Work} & \textbf{Year} & \textbf{Threat Model} & \textbf{Category} & \textbf{Defense} & \textbf{Metrics} & \textbf{Defense Overheads Presented} \vspace{5pt} \\
      \hline 
      \cite{Rajendran2013} & 2013 & I & Proximity Perturbation &Pin Swapping & Hamming Distance & \multicolumn{1}{c}{-*} \\
      \cite{Imeson2013} & 2013 & II & Wire Lifting &Wire Lifting & k-Distance & Power, Area, Delay and Wire-Length \\
      \cite{Vaidyanathan2014a} & 2014 &  I & Layout Obfuscation & Layout Obfuscation for SRAMs and Analog IPs & - & Performance, Power and Area \\
     \cite{Jagasivamani2014} & 2014 & I & Layout Obfuscation & Obfuscation Techniques & Neighbor Connectedness and Entropy & Performance and Area \\
     \cite{Otero2015} & 2015 & I & Layout Obfuscation & Automatic Obfuscation Cell Layout & Neighbor Connectedness and Entropy & Performance, Power and Area \\
     \cite{Xiao2015} & 2015 & I & Layout Obfuscation & Obfuscated Built-in Self-Authentication & Obfuscation Connection & Number of Nets \\
     \cite{Magana2016} & 2016 & I & Wire Lifting & Artificial Blockage Insertion & Number of Pins & \multicolumn{1}{c}{-*} \\
     \cite{Yang2016} & 2016 & I & Wire Lifting & Net Partition, Cell Hidden and Pin Shaken & - &  \multicolumn{1}{c}{-*} \\
     \cite{Wang2017} & 2017 & I & Proximity Perturbation & Routing Perturbation & Hamming Distance & Performance and Wire-Length \\
     \cite{Feng2017} & 2017 & I & Wire Lifting & Secure Routing Perturbation for Manufacturability & Hamming Distance & Performance and Wire-Length \\
     \cite{Sengupta2017} & 2017 & I & Proximity Perturbation & placement-centric Techniques & CCR & Performance, Power and Area \\
     \cite{Chen2017} & 2017 & II & Proximity Perturbation & Gate Swapping and Wire Lifting & Effective Mapped Set Ratio and Average Mapped Set Pruning Ratio & Wire-Length \\
     \cite{Patnaik2018} & 2018 & I & Wire Lifting & Concerted Wire Lifting & Hamming Distance & Performance, Power and Area \\
     \cite{Wang2018} & 2018 & I & Proximity Perturbation & Secure Driven Placement Perturbation & Hamming Distance & Power and Wire-Length\\
     \cite{Patnaik2018a} & 2018 & I & Proximity Perturbation & placement and routing perturbation & Hamming Distance & Performance, Power and Area \\
     \cite{Masoud2019} & 2019 & I & Layout Obfuscation & Isomorphic replacement for Cell Obfuscation & Isomorphic Entropy & \multicolumn{1}{c}{-*} \\
     \cite{Li2019a} & 2019 & II & Layout Obfuscation & Dummy Cell and Wire Insertion & k-security & Area and Wire-Length \\
    \hline \hline
      \multicolumn{7}{l}{\cellcolor{white}* Authors do not present any discussion regarding overhead.} 
    \end{tabular}
    \label{tab:sm_defenses}
\end{table*}

\section{Split Manufacturing Defenses} \label{sec:defenses}

Attacks toward Split Manufacturing showed promising results, as described in the previous section. A malicious attacker has the real potential to recover the missing BEOL connections. If the missing connections are successfully recovered, the security introduced by applying the technique is nullified. Therefore, straightforward Split Manufacturing is questioned by several works. Several authors proposed defense techniques that augment the technique, i.e., techniques that, when used together with Split Manufacturing, do increase the achieved security against attacks. In Table~\ref{tab:sm_defenses}, we compile a comprehensive list of defense techniques found in the literature. Each defense technique utilizes a different metric and Threat model, depending upon the type of attack they are trying to overcome. Since many of the studied defense techniques often introduce heavy PPA overheads, Table~\ref{tab:sm_defenses} also shows if the studied work assessed overheads and which ones were addressed. 

In the text that follows, the many defense techniques are divided into categories, namely Proximity Perturbation (i.e., change the location of cells or pins), Wire Lifting (i.e., move routing wires to upper layers), and Layout Obfuscation (i.e., hide the circuit structure). We present the categories in this exact order. For some techniques, it is worth mentioning that overlaps do exist and that techniques could be categorized differently. Thus, this categorization is our interpretation of the state of the art and may not be definitive. Furthermore, the boundaries between categories are not strict. For example, a technique may perform a layout modification that promotes proximity perturbation and leads to (indirect) wire lifting. In Tables~\ref{tab:def_res} and \ref{tab:def_res2}, we compile the results for the Proximity Perturbation and Wire Lifting categories, respectively. To demonstrate the effectiveness of each defense technique, we compile the results for when the attack is done with and without the defense. The results showed in Tables~\ref{tab:def_res} and \ref{tab:def_res2} are for the smallest and largest circuits addressed in each  studied work. Additionally, we show the PPA overhead introduced and, if specified, the split layer.
 
 \subsection{Proximity Perturbation}
 
Attacks toward split circuits are generally based on leveraging proximity information. The first category of defenses, Proximity Perturbation, addresses this hint left by the EDA tools. The goal of the techniques within this category is to promote changes in the circuit such that the proximity information between the FEOL pins is less evident. Therefore, the success rate of the proximity attacks is decreased.
 
In \cite{Rajendran2013}, the authors proposed pin swapping to overcome proximity attacks. Rearranging the partition pins can alter their distance in such a way that the attacker is deceived. As an example, if the pins $P_{G3,B,in}$ and $P_{G6,A,in}$ (Figure~\ref{fig:cir_part}) are swapped, the proximity attack will incorrectly guess the connection between $P_{G2,A,out}$ and $P_{G3,B,in}$. Thus, a sufficient number of pins have to be swapped in order to create a netlist that is significantly different from the original netlist (based on some sort of metric for similarity). In \cite{Rajendran2013}, Hamming distance is proposed as a way to quantify the difference between the outputs of the original netlist and the modified netlist. Assuming the outputs of a circuit are arranged as a string of bits, Hamming distance is defined as the number of bits that change when two instances of this string are compared to one another. The authors argued that the optimum netlist is created when the Hamming distance is 50\%. Therefore, inducing the maximal ambiguity for a potential attacker. Since the best rearrangement for N pins of partitions might take $N!$ computations (rather computationally expensive), pair-wise swapping of pins is considered in \cite{Rajendran2013}. Pair-wise swapping of pins results in $O(N^2)$ computations. 
   
The modified netlist is created based on a series of rules. Similarly, to the proximity attack, a list of candidates pins to be swapped is created before the actual swap is applied. Since not every pin can be swapped, a candidate pin to be swapped should:
   
   \begin{itemize}
       \item be an output pin of the partition where the target pin resides
       \item not be connected to the partition where the candidate pin resides
       \item not form a combinational loop
   \end{itemize}
   
Using the above constraints, a candidate pin is selected. The target pin also needs to be chosen carefully. In \cite{Rajendran2013}, IC testing principles \cite{testBook} and hints from the original proximity attack are used to choose the target pin. The swapping procedure is described in Algorithm \ref{alg:pin_swap}, where $TestMetric$ is a metric based on IC testing principles, such as stuck-at fault models which are still utilized in Test today. More details can be obtained from \cite{Rajendran2013}. The proposed defense technique is validated using ISCAS\textquotesingle85 circuits and the original proximity attack. For the smallest circuit, c17, it takes only one swap to achieve a Hamming Distance of 50\%. For the largest studied circuit, c7552, it takes 49 swaps. These results are summarized in Table~\ref{tab:def_res}.

As demonstrated in \cite{Rajendran2013}, rearranging the partition pins can thwart proximity attacks. However, according to Chen \etal \cite{Chen2017}, pin swapping at partition level has limited efficacy. They demonstrated that an attacker holding the FEOL layout as well as the nestlist can insert hardware trojans even when the defense approach of \cite{Rajendran2013} is applied. It must be highlighted that \cite{Chen2017} assumes threat model II, which we have previously argued that has the potential to nullify the vast majority of defenses towards split circuits. Thus, they proposed a defense to counter the threat from hardware trojans. Their defense incorporates the global wire-length information, with the goal to hide the gates from their candidate locations, and as result decreasing the effective mapped set ratio (EMSR). The EMSR metric is an attempt to quantify the ratio of real gates location of a given mapping during a simulated annealing-based attack. This defense consists of two steps, first a greedy gate swapping defense \cite{Wang2018}, and second, a measurement of the security elevation in terms of EMSR. The technique is evaluated using ISCAS\textquotesingle85 benchmarks circuits and the EMSR metric to quantify the defense effectiveness. The results are shown in Table~\ref{tab:def_res}.
   
Following the same principle of increasing the Hamming Distance, Wang \etal \cite{Wang2017} proposed a routing perturbation based defense. The optimum Hamming distance is sought to be achieved by layer elevation, routing detours, and wire decoys, while test principles are used to drive the perturbation choices. Layer elevation is essentially a wire lifting technique: without changing the choice of split metal layer, wires are forced to route using higher metal layers, thus being lifted from the FEOL to the BEOL. Intentional routing detours are a way to increase the distance between disconnected pins of the FEOL. If done properly, disconnected pins will not be the closest to each other, deceiving the proximity attack. In some cases, routing detours will increase the distance between disconnected pins, however, they still remain the closest to each other. In this scenario, wire decoys can be drawn near disconnected pins, in such a way that decoys are now the closest and will instead be picked as the ideal candidate pin. 

The perturbations proposed in \cite{Wang2017} can incur heavy overheads, and, for this reason, wires to be perturbed are chosen by utilizing IC test principles. In \cite{Wang2017}, fault observability, as defined in SCOAP \cite{SOAP}, is used as a surrogate metric for this task. The technique is evaluated using ISCAS\textquotesingle85 and ITC\textquotesingle99 benchmark circuits. For all studied circuits, the Hamming distance increased by an average of 27\% at a cost of only 2.9\% wire length overhead (WLO), on average. The results for the largest and smallest studied circuits are shown in Table~\ref{tab:def_res}.

   \begin{algorithm}[htb]
\DontPrintSemicolon
  
  \KwInput{Partitions}
  \KwOutput{List of target and swapping pins}
  $ListofTargetPins$ = $\emptyset$;\\
  $ListofSwappingPins$ = $\emptyset$;\\
  $ListofUntouchedPins$ = All partition pins and I/O ports;\\
   \While{Untouched output partitions pins or input ports exist}
   {
    
    \For{$UntouchedPin$}
        {
            $SwappingPins$ = \\
            BuildSwappingPinsList($UntouchedPin$);
            \For{$SwappingPin$ $\in$ $SwappingPins$}
                {
                    Compute \\
                    $TestMetric$($UntouchedPin,
                    SwappingPin$);
                }
        }
        Find the $TargetPin$ and $SwappingPin$ with the Highest $TestMetric$ from its SwappingPins;\\
        $ListofTargetPins += TargetPins$;\\
        $ListofSwappingPins += SwappingPins$;\\
        $ListofUntouchedPins -= TargetPins$;\\
        $LisofUntouchedPins -= SwappingPin$; \\
        Swap TargetPin and SwappingPin; \\
        Update netlist;
   }
   \textbf{Return:} ListofTargetPins and ListofswappingPins;
   \textbf{BuildSwappingPinList}($TargetPin$); \\
    \KwInput{$TargetPin P_{x,i,out}$}
  \KwOutput{$SwappingPins$ for $TagetPin$}
  \For{$Pin_J \in SwappingPins$}
    {
        \If{$CombinationalLoop(TargetPin,Pin_J)$}
        {
            $SwappingPins -= Pin_J$;
        }
    }
    \textbf{Return:} $SwappingPins$;
\caption{ Fault analysis-based swapping of pins to thwart proximity attack (adapted from \cite{Rajendran2013}).}
\label{alg:pin_swap}
\end{algorithm}
   
Sengupta \etal \cite{Sengupta2017} take a different direction from other works. They utilized an information-theoretic metric to increase the resilience of a layout against proximity attacks. As demonstrated in \cite{Sengupta2017}, mutual information (MI) can be used to quantify the amount of information revealed by the connectivity distance between cells. Mutual information is calculated by taking into account the cells connectivity $D$, if they are connected or not, and their Manhattan distance $X$, described by Eq. \ref{eq:mutual_info}, where $H[\cdot]$ is the entropy. The Manhattan distance of two cells is defined as the sum of horizontal and vertical distances between them. Entropy is a measure of disorder of a system. Therefore, in this work, entropy is utilized as a measure of disorder in the FEOL layer. The distribution of the variable $X$ for a given layout is determined pair-wise for all gates, allowing a straightforward computation of $I(X;D)$. Thus, layouts with the lowest mutual information, i.e., the correlation between cell connectivity and their distance is low, are more resilient against proximity attacks.

   \begin{equation}\label{eq:mutual_info}
       MI = I(X;D) = H[X]-H[X/D]
   \end{equation}
   
In order to minimize the information ``leaked'' from mutual information, \cite{Sengupta2017} applies cell placement randomization and three other techniques: g-color, g-type1, and g-type2. Randomizing the cell placement can achieve the desired low mutual information; however, the PPA overhead incurred is excessive. Minimizing mutual information without excessive PPA overhead can be achieved by the other techniques. From a graph representation of the circuit, graph coloring can be used to hide connectivity information, where gates of the same color must not be connected. Thus, the resulting colored netlist is then partitioned by clustering all cells of same color together. During cell placement, the cells with the same color will be confined within their respective clusters. According to \cite{Sengupta2017}, these constraints naturally mitigate the information leakage to a great extent. The g-color technique utilizes only the graph coloring as described above. The other two, g-type1 and g-type2, consider the type of the gate when creating clusters. The g-type1 approach clusters gates only by their functionality, while g-type2 utilizes functionality and the number of inputs for clustering. The authors assessed their techniques utilizing ISCAS\textquotesingle85 and MNCN benchmark suites. Results for the smallest and largest circuits are shown in Table~\ref{tab:def_res}.

\begin{table*}[htb]
\rowcolors{2}{gray!25}{white}
    \centering
     \caption{Results for Defense Techniques based on Proximity Perturbation.}
    \begin{tabular}{p{.8cm}p{1cm}p{1.2cm}p{2.5cm}lp{2.2cm}p{.5cm}p{1.5cm}p{1.5cm}}
    \hline \hline \\ [-1.5ex]
       \textbf{Work} & \textbf{Attack Type} & \textbf{Benchmark} & \textbf{Defense Technique} & \textbf{Defense Metric} & \textbf{Defense Overhead} & \textbf{Split Layer} &  \textbf{Result without Defense} & \textbf{Result with Defense} \vspace{5pt} \\
       \hline
       \cite{Rajendran2013} & Proximity & c17   & - & Hamming Distance   & 1 Swap for 50\% HD      & -*  & 100\% CCR  & 78\% CCR\\
       \cite{Rajendran2013} & Proximity & c7552 & - & Hamming Distance   & 49 Swaps for 50\% HD    & -*  & 94\% CCR   & 91\% CCR\\
       \cite{Chen2017}      & Proximity & c432  &  Modifed Greedy Gate Swapping & EMSR & 75\% of WLO & -* & 90\% EMSR & 25\% EMSR \\
       \cite{Chen2017}      & Proximity & c432  &  Modifed Greedy Gate Swapping & EMSR & 300\% of WLO & -* & 78\% EMSR & 10\% EMSR \\
       \cite{Wang2017}      & Proximity & c432  & - & Hamming Distance   & 3.1\% WLO for 46.1\% HD & -*  & 92.4\% CCR & 78.8\% CCR \\
       \cite{Wang2017}      & Proximity & c432  & - & Hamming Distance   & 4.1\% WLO for 31.7\% HD & -*  & 62.8\% CCR & 37.9\% CCR \\
       \cite{Sengupta2017}  & Proximity & c432  & Random & Mutual Information & < 10\% PPA               & M1 & 17\% CCR   & < 1\% CCR \\
       \cite{Sengupta2017}  & Proximity & c432  & g-color & Mutual Information & < 10\% PPA              & M1 & 17\% CCR   & 2\% CCR \\
       \cite{Sengupta2017}  & Proximity & c432  & g-type1 & Mutual Information & < 10\% PPA              & M1 & 17\% CCR   & 6\% CCR \\
       \cite{Sengupta2017}  & Proximity & c432  & g-type2 & Mutual Information & < 10\% PPA              & M1 & 17\% CCR   & 4.5\% CCR \\
       \cite{Sengupta2017}  & Proximity & c7552  & Random & Mutual Information & < 10\% PPA               & M1 & 13\% CCR   & < 1\% CCR \\
       \cite{Sengupta2017}  & Proximity & c7552  & g-color & Mutual Information & < 10\% PPA              & M1 & 13\% CCR   & 2\% CCR \\
       \cite{Sengupta2017}  & Proximity & c7552  & g-type1 & Mutual Information & < 10\% PPA              & M1 & 13\% CCR   & 4\% CCR \\
       \cite{Sengupta2017}  & Proximity & c7552  & g-type2 & Mutual Information & < 10\% PPA              & M1 & 13\% CCR   & 3\% CCR \\
       \cite{Wang2018}      & SAT       & c432   & BEOL+Physical  & Perturbation & 4.5\% WLO              & -*  & 58\% CCR   & 56\% CCR \\
       \cite{Wang2018}      & SAT       & c432   & Logic+Physical & Perturbation & 5.57\% WLO             & -*  & 58\% CCR   & 58\% CCR \\
       \cite{Wang2018}      & SAT       & c432   & Logic+Logic    & WLD          & 1.68\% WLO             & -*  & 58\% CCR   & 52\% CCR \\
       \cite{Wang2018}      & SAT       & b18    & BEOL+Physical  & Perturbation & 8.06\% WLO             & -*  & 15\% CCR   & 14\% CCR \\
       \cite{Wang2018}      & SAT       & b18    & Logic+Physical & Perturbation & 1.70\% WLO             & -*  & 15\% CCR   & 17\% CCR** \\
       \cite{Wang2018}      & SAT       & b18    & Logic+Logic    & WLD          & 0.61\% WLO             & -*  & 15\% CCR   & 16\% CCR** \\
       \cite{Patnaik2018a} & Proximity  & c432   & Netlist Randomization & Hamming Distance & < 10\% PPA overall & -* & 92.4\% CCR  & 0\% CCR \\
       \cite{Patnaik2018a} & Proximity  & c7552  & Netlist Randomization & Hamming Distance & < 10\% PPA overall & -* & 94.4\% CCR  & 0\% CCR \\
       \hline
       \hline
        \multicolumn{9}{l}{\cellcolor{white}* Split layer not specified by the authors.} \\
        \multicolumn{9}{l}{\cellcolor{white}** These results are counter-intuitive, the applied defense degrades the metric.} 
    \end{tabular}
   
    \label{tab:def_res}
\end{table*}   
 
Similar to the pin swapping technique proposed in \cite{Rajendran2013}, Wang \etal \cite{Wang2018} proposed a placement-based defense with the same objective of deceiving a proximity attack by perturbing proximity information. Differently from pin swapping, their placement-based defense considers the incurred wire-length overhead as a metric. This technique is based on changing gate locations such that the proximity hint is no longer effective. Their algorithm consists of two phases, one to select which gates to be perturbed and a second phase where the selected gates are (re)placed. Gate selection is done by extracting a set of trees using two techniques, BEOL-driven and logic-ware extraction. The first approach selects all gate trees that contain any metal wires in the FEOL, i.e., connections that are not hidden from the attacker. The second approach considers the wire-length impact and the gate tree impact on the overall security. After extracting the set of trees, the placement perturbation is done in one of two ways: physical-driven or logic-driven. For each extracted tree, the physical-driven perturbation changes the location of gates using a Pareto optimization approach. Also, each solution is evaluated by its wire-length overhead and a perturbation metric that discerns the placement difference from the original layout. According to \cite{Wang2018}, geometric-based difference alone may be insufficient to enhance the split circuit security. Thus, a logic-driven perturbation is performed with a weighted logical difference (WLD) metric, which encourages perturbation solutions with large logical difference from its neighbors. The authors assessed their techniques combining the gate selection and perturbation as BEOL+Physical, Logic+Physical and Logic+Logic, using ISCAS\textquotesingle85 and ITC\textquotesingle99 circuit benchmarks. Results for the smallest and largest circuits considered are shown in Table~\ref{tab:def_res}.
  
A considerably different approach is proposed by Patnaik \etal \cite{Patnaik2018a}, whereas netlist modifications are promoted (instead of placement/routing modifications during physical synthesis). The goal is to modify the netlist of a design in order to insert (partial) randomization. According to \cite{Patnaik2018a}, this approach helps to retain the misleading modifications throughout any regular design flow, thereby obtaining more resilient FEOL layouts where the netlist changes are later ``corrected'' in the BEOL. This methodology is implemented as an extension to commercial EDA tools with custom in-house scripts. The process goes as follows: first, the netlist is randomized. Second, the modified netlist is place and routed. Lastly, the true functionality is restored by re-routing in the BEOL. For the netlist randomization, pairs of drivers and their sinks are randomly selected and swapped. This is done in such way to avoid combinational loops that may be introduced by swapping. The modified netlist then is place and routed, utilizing a `do not touch'\footnote{This terminology is used in IC design to mean that a specific cell or family of cells should not be optimized, i.e., not to be touched.} setting for the swapped drivers/sinks to avoid logic restructuring/removal of the related nets. Finally, the true connectivity is restored in the BEOL with the help of correction cells \cite{Patnaik2018a} that resemble switch boxes. The technique is evaluated using ISCAS\textquotesingle85 circuits, and the results for the largest and smallest circuit are shown in Table~\ref{tab:def_res}.
  
 \subsection{Wire Lifting}
 
Hiding routing information from untrusted foundries is the main objective of the Split Manufacturing technique. Since attacks mainly rely on hints left by EDA tools to recover the missing BEOL connections, the amount of hidden information is related to the circuit performance -- splitting the circuits at low metal layers increases the security level. Following the same idea, wire lifting proposes `lifting' wires from the FEOL layer to the BEOL. That is, changing the routing to split metal layers has the potential to increase the security level. 

  \begin{table*}[htb]
    \rowcolors{2}{gray!25}{white}
    \centering
     \caption{Results for Defense Techniques based on Wire Lifting.}
    \begin{tabular}{cp{1cm}p{1.5cm}p{2.5cm}llp{1cm}p{1.5cm}p{1.5cm}}
    \hline \hline \\ [-1.5ex]
       \textbf{Work} & \textbf{Attack Type} & \textbf{Benchmark} & \textbf{Defense Technique} & \textbf{Defense Metric} & \textbf{Defense Overhead} & \textbf{Split Layer} & \textbf{Result without Defense} & \textbf{Result with Defense} \vspace{5pt}\\
       \hline
    \cite{Imeson2013} & SAT & c432 & Wire Lifting & \textit{k-security} & 477\% of WLO & -* & k=1 & k=48 \\
    \cite{Magana2016} & Proximity & Superblue 1 & Routing Blockage Insertion & $E[LS]$ & Not Presented & M4 &  1.51 &  1.77 \\
    \cite{Magana2016} & Proximity & Superblue 1 & Routing Blockage Insertion & $FOM$ & Not Presented & M4 & 1222.8 &  1433 \\
    \cite{Patnaik2018} & Proximity & c432 & Concerted Lifting & Hamming Distance & 7.7\% of Area & Average** & 23.4 & 45.9 \\
    \cite{Patnaik2018} & Proximity & c432 & Concerted Lifting & CCR & 13.2\% of Power & Average** & 92.4 & 0 \\
    \cite{Patnaik2018} & Proximity & c7552 & Concerted Lifting & Hamming Distance & 16.7\% of Area & Average** & 1.6 & 25.7 \\
    \cite{Patnaik2018} & Proximity & c7552 & Concerted Lifting & CCR & 9.3\% of Power & Average** & 97.8 & 0 \\
    \cite{Feng2017} & Proximity & c2670 & CMP-Friendly & Hamming Distance & 3.4\% of WLO & -* & 14.5\% & 20.4\% \\
    \cite{Feng2017} & Proximity & c2670 & CMP-Friendly & CCR(\%) & 3.4\% of WLO & -* & 48.1\% & 33.4\% \\
    \cite{Feng2017} & Proximity & b18 & CMP-Friendly & Hamming Distance & 0.4\% of WLO & -* & 21.6\% & 27.6\% \\
    \cite{Feng2017} & Proximity & b18 & CMP-Friendly & CCR(\%) & 0.4\% of WLO & -* & 12.1\% & 10.7\% \\
    \cite{Feng2017} & Proximity & c2670 & SADP-Compliant & Hamming Distance & 7.49\% of WLO & -* & 14.5\% & 24.4\% \\
    \cite{Feng2017} & Proximity & c2670 & SADP-Compliant & CCR(\%) & 7.49\% of WLO & -* & 48.1\% & 6.4\% \\
    \cite{Feng2017} & Proximity & b18 & SADP-Compliant & Hamming Distance & 4.64\% of WLO & -* & 21.6\% & 29.6\% \\
    \cite{Feng2017} & Proximity & b18 & SADP-Compliant & CCR(\%) & 4.64\% of WLO & -* & 12.1\% & 2.7\% \\ \hline
    \cite{Yang2016} & Proximity &  s526 & Net Partitioning & CCR(\%) & Not Presented & -* & 40\%*** & 0\%*** \\
    \cite{Yang2016} & Proximity &  s526 & Net Partitioning \& Cell Hiding & CCR(\%) & Not Presented & -* & 40\%*** & 0\%*** \\
    
     \cite{Yang2016} & Proximity &  s526 & Net Partitioning \& Cell Hiding \& Pin Shaking & CCR(\%) & Not Presented & -* & 40\%*** & 0\%*** \\
    \cite{Yang2016} & Proximity &  s9234.1 & Net Partitioning & CCR(\%) & Not Presented & -* & 30\%*** & 4\%*** \\
    \cite{Yang2016} & Proximity &  s9234.1 & Net Partitioning \& Cell Hiding & CCR(\%) & Not Presented & -* & 30\%*** & 1.5\%*** \\
     \cite{Yang2016} & Proximity &  s9234.1 & Net Partitioning \& Cell Hiding \& Pin Shaking & CCR(\%) & Not Presented & -* & 30\%***& 1.5\%*** \\
         \hline    
         \hline
         \multicolumn{9}{l}{\cellcolor{white}* Split layer not specified by the authors.} \\
          \multicolumn{9}{l}{\cellcolor{white}** Results are given as an average between M3, M4, and M5.} \\
          \multicolumn{9}{l}{\cellcolor{white}*** These results cannot be directly compared with previous ones as the transistor technology is vastly different.} 
         \end{tabular}
    \label{tab:def_res2}
\end{table*}

Wire lifting was first presented by Imerson \etal \cite{Imeson2013} where Split Manufacturing is considered as a 3D IC implementation \cite{3DwhitePaper}. For the sake of argument, we will continue to refer to this technique as Split Manufacturing, even if the notion of untrusted FEOL vs. trusted BEOL is shifted. This type of 3D implementation consists of two or more independently manufactured ICs, where each IC represents a tier that is vertically integrated on top of each other. Connections between the tiers are done using vertical metal pillars, referred to as through-silicon vias (TSVs). In \cite{Imeson2013}, a 3D implementation consisting of two tiers is used for their experiments. The bottom tier containing the transistors and some routing wires (akin to the FEOL), and the top tier, containing only routing wires (akin to the BEOL). Regarding the manufacturing of these 3D ICs, the bottom tier is built in a high-end untrusted foundry, and the top tier is built in an also untrusted foundry (not necessarily high-end, however).
     
In \cite{Imeson2013}, threat model II is used, i.e., the adversary is assumed to possess the entire netlist. The problem is formulated as the attacker being the FEOL foundry, which in turn also possesses the so called `unlifted netlist' extracted from the FEOL layout. By utilizing a graph to represent the circuits as previously described, the attacker seeks a bijective mapping of gates of the unlifted netlist to gates in the complete netlist. According to \cite{Imeson2013}, if the attacker can distinguish any gate between the two netlists, the split circuit does not provide any security. A security notion is provided by the authors, based on existing multiple mapping between gates in the unlifted and complete netlists. Referred to as \textit{k-security}, this metric qualifies that gates across the design are indistinguishable from at least $k-1$ other gates. Thus, a defender wants to lift wires in a way to guarantee the higher $k-security$ possible. Two procedures are proposed to achieve this goal, one utilizing a greedy heuristic targeted at small circuits (due to scalability issues), and another procedure that utilizes partitioning to solve those issues. For their experimental study, they have utilized the ISCAS\textquotesingle85 benchmark suite and a DES crypto circuit with approximated 35000 gates. The results are shown in Table~\ref{tab:def_res2}, where $k=1$ is the original circuit and $k=48$ is achieved when all the wires are lifted. It is worth to mention that; besides the notion of the security metric, their defense technique was not validated using an actual proximity attack towards the modified netlist.

An artificial routing blockage\footnote{This terminology is used in IC design to mean that a specific area should be avoided by the EDA tool for a specific task. A blockage can be for placement and/or for routing.} insertion that promotes wire lifting is proposed by Maga\~na \etal \cite{Magana2016}. The goal of this technique is to deceive proximity attacks by wire lifting. As discussed before, the objective of commercial EDA tools is to guarantee the best PPA possible. During the routing stage, lower metals are preferred for signal routing, promoting better PPA. Thus, routing blockages can be inserted at the split layer, forcing signals to be routed above the split layer. The result is an artificial wire lifting done during the routing stage. 
     
Applying this type of procedure must be done considering the design routability and overhead introduced, as well as top level floorplan decisions for the power grid, clock distribution, and resources for busses. Larger designs are generally difficult to be routed -- simply reducing the number of routing layers can make the design unroutable. In \cite{Magana2016}, a procedure is proposed to insert routing blockages ensuring the design routability is kept. After a first routing stage, the design is divided into small rectangular non-overlapping windows. The routing congestion is then analyzed in each window at the split layer for the blockage insertion. If the area has capacity for more routing, a routing blockage is inserted, otherwise the original routing is kept. Utilizing ISPD\textquotesingle11 circuits, the technique is evaluated using the proximity attack proposed by \cite{Magana2016}, and its effectiveness is measured using two metrics, $E[LS]$ and $FOM$. The $E[LS]$ metric reports the candidate list size, being an average over different search areas. The $FOM$ metric is a figure of merit obtained from the ratio of candidates list size divided by the search area, when averaged over all the search areas at the split layer. According to \cite{Magana2016}, a higher value of $FOM$ means it is more challenging for an attack to be mounted because of the density of candidates (over the same search area). The results for the \textit{Superblue 1} circuit are shown in Table~\ref{tab:def_res2}.
     
Design for Manufacturability (DFM) has become an extremely important aspect of IC design for many years now. Manufacturing an IC is a sensitive process that involves many critical steps. Hence, a layout is required to be compliant to several rules to ensure its manufacturability. A layout is said to be manufacturable if there are no DRC violations. However, for a design to also achieve high yield, the layout must also pass strict DFM checks. The most common checks are related to wire and via density over predetermined region sizes. Until now, defense techniques discussed were mainly concerned about security and PPA overheads. Feng \etal \cite{Feng2017} argued that previous works have largely neglected manufacturability concerns. Therefore, they proposed two wire-lifting techniques that address two important DFM-related techniques: Chemical Mechanical Planarization (CMP) and Self-Aligned Double Patterning (SADP) \cite{DFM1}. The first technique, CMP-friendly routing defense is divided into layer elevation, wire selection, and re-routing. Layer elevation selects wires for lifting according to following principles \cite{Feng2017}:
      
      \begin{itemize}
          \item The wire has a significant logic difference from its neighboring wires. As such, an incorrect connection in attacking this wire may lead to more signal differences.
          \item The wire has large observability such that an erroneous guess by the adversary can easily affect the circuit primary output signals.
          \item The wire segment is originally at a wire-dense region. The wire density of this region would be reduced by the layer elevation and makes the corresponding FEOL layer have more uniform wire density.
          \item The BEOL region where the wire is elevated to has low wire density so that the density of the corresponding BEOL layer is more uniform.
      \end{itemize}
      
Principles 1 and 2 have the goal to increase security in the same way as described in \cite{Wang2017}. After the wire lifting step, a set of wires is selected for re-routing. The selection has two purposes, CMP-friendliness and security improvement. For CMP-friendliness, wires located in dense regions are selected to be re-rerouted in sparse areas. For the security improvement, decoys are inserted if the routing detour passes through a sparse area. A suspicious attacker may realize that the detour is a defense measure. After selecting the set of wires to be re-routed, wires are re-routed one at a time. According to \cite{Feng2017}, their routing approach considers wire density, while the routing perturbation proposed by \cite{Wang2017} can be solely focused on security, and may not be CMP-friendly. Utilizing a graph representation, their re-routing method is based on the Dijkstra's shortest path algorithm \cite{Djktas} where the density of wires is used as a metric. 
    
With a few exceptions, the SADP-compliant routing defense follows the same approach as described above. During wire lifting, the density is not considered. Wire re-routing is actually wire extension of FEOL wires as in \cite{SADPP}. This wire extension of FEOL wires inevitably leads to re-routing of connected BEOL wires. According to \cite{Feng2017}, solving SADP violations by wire extension can also increase security, as its increase the distance between vias. The wire extension for simultaneous SADP-compliance and security is realized using Integer Linear Programming. In their experiments, ISCAS\textquotesingle85 and ISPD\textquotesingle11 are used to evaluate their techniques. Each technique, CMP-friendly and SADP-compliant routing, is evaluated separately. The results for the smallest and largest circuits are shown in Table~\ref{tab:def_res2}.

Wire lifting approaches, in general, are not cost-free. As shown in the discussed results, wire-lifting based defenses introduce a considerable PPA overhead. An approach to establish a cost-security trade-off is proposed by Paitinak \etal \cite{Patnaik2018}, i.e., PPA margins for a given security budget. In \cite{Patnaik2018}, a concerted wire-lifting method is proposed. The authors claim to enable higher degrees of security while being cost-effective. For their method, custom elevating cells are used for executing the wire-lifting. Elevating cells connect gates or short wires directly to the first layer above the split layer. Their wire-lifting method utilizes three strategies: lifting high-fanout nets, controlling the distance for open pin pairs, and obfuscation of short nets. High-fanout nets are chosen to be lifted for two reasons: (a) a wrong connection made by the attacker propagates the error to multiple locations, and, (b) introduces multiple open pin pairs. As the attack to overcome is the proximity one, controlling the distance between open pin pairs is necessary, which is achieved at will simply by controlling the placement of the elevating cells. According to \cite{Patnaik2018}, short nets may be easy for an attacker to identify and localize (from assessing driving strengths). Short wires are obfuscated by inserting an elevating cell with two pins close to each other, one being the true connection and the other a dummy connection. Finally, wires are lifted according to those strategies until a given PPA budget is reached. For their experimental study, ISCAS\textquotesingle85 and ISPD\textquotesingle11 circuits are utilized. However, results for attacks are presented only for ISCAS\textquotesingle85 circuits. For ISPD\textquotesingle11, only the PPA impact result introduced by their technique is presented. Once again, we present the results for the smallest and largest of the studied circuits in Table~\ref{tab:def_res2}.
    
While the majority of studies reported in our survey make use of conventional transistors (bulk CMOS technologies with either planar or FinFET transistors), Yang \etal \cite{Yang2016} proposed a design methodology to secure Split Manufacturing for Vertical Slit Field Effect Transistor (VeSFET)-based integrated circuits. VeSFET is a twin-gate device with a horizontal channel and four metal pillars implementing vertical terminals \cite{vesfet1}. While a detailed explanation on VeSFETs is beyond the scope of this work, we do highlight the differences between VesFETs and conventional transistors. In contrast with conventional transistors, a VeSFET can be accessed by both top and bottom pillars, allowing two-side routing and offering a friendly monolithic 3D integration \cite{vesfet1, vesfet2, vesfet3}. While we have so far considered ICs that have two distinct layers, the FEOL and BEOL, a VeSFET-based IC has tiers of the layer containing the transistors. Connections between tiers can be made directly, same as TSV by the pillars, or by a layer containing connections between tiers. A 2D VeSFET design contains only one tier and both top and bottom connections, whereas a 3D design contains two or more tiers. In summary, the notion of tier is pushed down to the transistor level in this device topology, thus making it an interesting platform for Split Manufacturing.
    
The method proposed by \cite{Yang2016} assumes that both foundries are untrusted and have the same capability (i.e., same technology). For 2D designs, the first foundry manufactures the tier with the top connections, comprising most of the connections. Then, the rest of the bottom connections, comprising of the critical minority connections, are completed by the second foundry. For 3D IC designs, they proposed special types of standard cells, referred as cell A and B. Cell A has two tiers that are visible and manufactured by the first foundry, as well as inter-tier connections. Cell B has only the top tier visible and manufactured by the first foundry, the low tier is completed by the second foundry, without inter-tier connections. Thus, transistors can be hidden from the first foundry as a security feature. Vulnerabilities claimed by \cite{Yang2016} for both 2D and 3D methods are described in Table~\ref{tab:vesfet_vul}. Practices of reverse engineering and IC overbuilding are claimed to be impossible because the first foundry controls the number of wafers available to the second foundry.
   
Increasing the security of both 2D and 3D VesFET designs is achieved by net partitioning, and exclusively for 3D designs, by transistor hiding and pin shacking. Net partitioning is performed similarly to the wire lifting techniques described above, where nets are chosen to be routed in the bottom connection layer, thus, hiding those from the first foundry. Their selection method is done by selecting nets from sequential logic. First, all the high-fanout nets are selected to be partitioned. Next, the remaining nets are selected by a search area, where two approaches are used, distance-first search and high-fanout first search. In distance-first method, a pin in a predefined search window connecting to an un-partitioned net is selected when it has the minimum distance to the currently processed pin pair. The FO-first search method selects the pin connecting to a net having the highest FO in the searching window. Transistor hiding in 3D designs is done by utilizing cells similar to the cell B. Cells connected only by partitioned nets are candidates for hiding. After selecting the candidates, availability of unused transistors that are accessible to the second foundry in the lower tiers of the nearby cells is checked. If the available transistor count is sufficient, then the cell is hidden. The empty space created could provide clues for the first foundry about the security technique. Pin shaking is then applied to obfuscate the empty spaces. Some nearby cells are moved to this area to obfuscate the layout for any distance-based proximity attackers. In \cite{Yang2016}, 10 MCNC LGSynth\textquotesingle91 benchmark circuits are used to evaluate the effectiveness of their methodology. The best and worst results are shown in Table~\ref{tab:def_res2}. It is worth to mention that, even though the VesFET implementation mimics the layered structure of Split Manufacturing, the results cannot be compared side by side in a fair manner.
   
    \begin{table}[htb]
    \rowcolors{2}{gray!25}{white}
       \centering
       \caption{Vulnerabilities of Split Manufactured VesFET-based Designs Described by \cite{Yang2016}.}
       \begin{tabular}{lll}
       \hline \hline \\ [-1.5ex]
          \textbf{Threats}  & \textbf{1st Foundry} & \textbf{2nd Foundry} \vspace{5pt} \\
        \hline
         \parbox{1cm}{Design\\ Reconstruction}   & \parbox{2.5cm}{2D IC: Very Difficult \\ 3D IC: Impossible} & \parbox{2cm}{Impossible due to a very limited\\ information} \\
         Trojan Insertion        & \parbox{2.5cm}{Possible, but will be\\ detected}  & \parbox{2cm}{No control of devices} \\
         \parbox{1cm}{Reverse\\ Engineering}      &  Meaningless & Impossible\\
         IC Overbuilding           & Meaningless  & Impossible \\
         \hline
         \hline
       \end{tabular}
       \label{tab:vesfet_vul}
   \end{table}
   
\subsection{Layout Obfuscation}

The main goal of Split Manufacturing -- to hide sensitive information from untrusted foundries -- is compromised once we start to consider more regular structures such as memory. Even without knowing where all the routing goes to, an attacker can easily identify regular structures just by looking at the FEOL layout, possibly leading to easier attacks. Mitigating attacks towards regular structures could be done by obfuscating those structures in such a way that they become indistinguishable. In this section, we discuss works that propose layout obfuscation techniques to be used in a Split Manufacturing context.
    
During the development of a modern IC, third-party IPs are sought to close a technological gap or to minimize time-to-market. IPs are typically categorized as soft and hard IPs: soft IPs typically come in code form, giving the customer flexibility to modify the IP such that it meets a given specification during synthesis. Therefore, soft IPs do not present a direct challenge for a Split Manufacturing design flow. Perhaps, and on a very specific scenario, a given IP can facilitate a proximity attack because it promotes certain library cells over others (i.e., it leads to a biased composition).

On the other hand, hard IPs are completely designed by the vendor and are technology dependent. In some instances, the vendor only provides an abstract of the IP; the customer then has to rely on the foundry to replace the abstract by the actual layout. Thus, splitting a hard IP is not trivial. Additional information is needed to be provided by the vendor, which is not guaranteed to be provided, making the IP completely incompatible with Split Manufacturing. Even when the customer holds the entirety of the IP layout, differences between the FEOL foundry and the BEOL foundry could make the IP no longer compliant and therefore virtually useless. Furthermore, defense techniques cannot be applied due to the lack of information or lack of feasibility. Hard IPs, such as embedded memories and specialized analog circuits, have been heavily optimized for maximum compactness, performance and yield. In today's IP market, there is still little concern with security in general, so it is not conceivable that any vendors will start to offer split IP any time soon.
    
The security of hard IPs in a Split Manufacturing context was first analyzed by Vaidyanathan \etal \cite{Vaidyanathan2014a}. A recognition attack flow was proposed for this purpose. An attacker holding the FEOL layer starts his attack by isolating a target embedded memory or analog hard IP. Since the targeted hard IP has a high probability of being constructed by compilation of leaf-cells, layout pattern recognition software \cite{swDegate} can be used for trivial leaf-cell identification. After recognizing all the leaf-cells, the attacker attempts to infer the missing BEOL connections. Using proximity hints together with the knowledge about the regularized structure, the connections have a high likelihood to be guessed correctly. Demonstrated in \cite{Vaidyanathan2014a}, embedded memories, such as SRAM, are susceptible to the proposed recognition attack. Defending against recognition attacks can be achieved by means of obfuscation. According to \cite{Vaidyanathan2014a}, SRAM IPs can be obfuscated by the following methods:
    
    \begin{itemize}
        \item Randomization of periphery cells, thus avoiding predictable connections. 
        \item Minimization of regularized topologies used for peripheral circuits such as pre-decoders, word line decoders, sense amplifiers, etc.
        \item Adding non-standard application-specific functions to improve obfuscation and performance.
    \end{itemize}
    
A synthesis framework is proposed by \cite{Vaidyanathan2014a} to obfuscate SRAM IPs. Referred as application-specific SRAM, the methodology synthesizes SRAMs using augmented bitcell arrays and standard cell logic IP (instead of using leaf-cells). Such synthesis, when compared with conventional SRAM compilation, accomplishes all the three obfuscation goals described above while still providing similar performance. 
    
Analog hard IPs are also vulnerable to recognition attacks. In contrast with embedded memories (that are often compiled), analog hard IPs are mostly hand designed to cater for a challenging specification or interface. Even when such degree of customization is employed, the majority of the design is done utilizing leaf-cells (e.g., current mirrors, matched arrays, etc.). Thus, disclosing important information that could be used as leverage for recognition attacks. In \cite{Vaidyanathan2014a}, two methods are proposed to defend analog hard IPs against such attacks:
    
    \begin{itemize}
        \item Obfuscation of analog leaf-cells.
        \item Use of diverse topologies and architectures that enable obfuscation and efficiency. 
    \end{itemize}
   
Next, let us discuss the techniques utilized in order to achieve the goals listed above. First, adding camouflaging dummy transistors in empty spaces can turn leaf-cells indistinguishable. Second, regularizing transistor widths, which allows transistor with different channel lengths to abut each other, thereby obscuring boundaries across different sized transistors. Third, utilizing the same idea behind wire-lifting, routing blockages can be inserted between transistors below the split layer. Such routing scheme would make it difficult to infer the missing BEOL connections, virtually in the same way as it does for a standard-cell based design. 
   
To demonstrate the feasibility and efficacy of their proposed approaches, the authors of \cite{Vaidyanathan2014a} designed and fabricated test chips in 130nm technology. For comparison, the same designs were Split Manufactured and conventionally manufactured. Split Manufacturing used Global Foundries Singapore as the untrusted foundry and IBM Burlington as the trusted foundry. Conventional manufacturing was entirely done in Global Foundries Singapore. The first reported design is a smart SRAM that targets an imaging application. Two implementations of a parallel 2x2 access 1Kb SRAM were demonstrated. For conventional manufacturing, the SRAMs were traditionally implemented, and for Split Manufacturing, the SRAMs were implemented using their smart synthesis approach. For their measurements, 10 chips were used to demonstrate the feasibility regarding PPA. Area reported for the split manufactured samples was 75\% of the conventional approach, and, while the power consumption was 88\%. Performance was the same between conventional and split manufactured, i.e., both could work with the same clock frequency. The PPA advantage of Split Manufacturing reported in \cite{Vaidyanathan2014a} is not from the manufacturing itself. This advantage is from their smart memory synthesis approach, that was not applied on the conventional manufacturing samples.

The second demonstrated design is a DAC with statistical element selection. The test chip contains a high resolution 15-bit current steering DAC. Only a description of the results is presented, where the authors claim there are tiny measurements differences between the performance of the conventional and the split manufactured, emphasizing that the differences are within measurement noise.
  
An attacker trying to reverse engineer a split IC will try to recover the maximum number of connections as possible, while minimizing the Time To Evaluate (TTE), i.e., the amount of time needed to reverse engineering the IC. For Jagasivamani \etal \cite{Jagasivamani2014}, the goal of a designer seeking to secure his design is to create an IC with a high TTE while being cost-effective regarding design effort and PPA overheads. If the TTE is high enough, an adversary would be discouraged from reverse engineering the IC. To achieve this goal, \cite{Jagasivamani2014} proposed obfuscation methods that do not require any modifications to standard cells nor the implementation of any specialized cell. 
   
Four techniques are proposed by \cite{Jagasivamani2014} for layout obfuscation, (1) limited standard-cell library, (2) smart-dummy cell insertion, (3) isomorphic cells and (4) non-optimal cell placement. Along with the techniques, a set of metrics is presented to help assess the obfuscation level of a design. \textit{Neighbor connectedness}, a measure of how interconnected cells are to their respective neighbors, i.e., how much proximity information is exposed to the attacker. For a specific cell, this metric is computed as how many connections that cell has for a given radius around it. \textit{Standard-cell composition bias}, a metric that addresses the effort required for composition analysis of a design. The bias signature could leave information of the function of the cell. Thus, this metric measure how skewed a design is according to a specific bias cell. In \cite{Jagasivamani2014}, they utilized three types of bias cells for this analysis: XOR-type, flip-flop type, and adder type of cells. \textit{Cell-level obfuscation}, a metric that measures the percentage of standard-cells that have been obfuscated. \textit{Entropy}, which is similar to the concept of mutual information previously discussed. 
   
Technique (1) aims to achieve obfuscation by reducing the use of complex cells and instead favor only simple cells to compose the design (i.e., to prefer single stage cells over complex multi-stage cells). Removing specialized complex cells could obfuscate functional signatures due to the larger granularity that is employed to construct the cell. However, since the functionality of complex gates will have to be reconstructed through basic cells, a heavy PPA overhead is likely to occur when applying this technique. Technique (2) aims to obfuscate composition analysis by adding dummy cells in such a way that a neutral bias composition is achieved. Dummy cells are inserted as spare cells \footnote{Spare cells are extra logic usually inserted during physical synthesis. These cells are used when an engineering change order (ECO) is required, such that small tweaks to the circuit logic can be performed with minimal changes to placement and routing.}, focusing solely on obfuscating the composition analysis. Technique (3) obfuscates the layout by regularizing the shapes of the cells in a library. All layouts of logic standard cells are made FEOL-identical such that the overall circuit layout appears to be a sea of gates. The functionality of the cells is defined later by the BEOL connections. Thus, the true functionally of the cell is hidden at the BEOL, making cell-level identification harder. Technique (4) employs the same strategy from placement perturbation discussed before. 
    
For their experimental study, the authors of \cite{Jagasivamani2014} made use of a multiplier block with a high number of adder cells and a crypto-like circuit. Experiments were separated into limited library and smart dummy insertion. Results are shown as a percentage relative to the baseline circuit, i.e., without any protection approach applied. Neighbor connectedness (\%) for a radius $\leq 25nm$ decreased substantially for both test cases and circuits (for more information see \cite{Jagasivamani2014}). Overhead results are shown in Table~\ref{tab:over_javs}, where the figures presented are normalized with respect to the baseline circuit.
    
    \begin{table}[htb]
        \centering
        \caption{Impact on Performance from the Defense Approaches of \cite{Jagasivamani2014}.}
        \begin{tabular}{cccc}
        \hline \hline \\ [-1.5ex]
           \textbf{Benchmark}  &  \textbf{Metric} & \textbf{Limited Library} & S\textbf{mart Dummy} \vspace{5pt}\\
           \hline
           \multirow{2}{*}{ mult24} &\cellcolor[gray]{0.9}  Area         & \cellcolor[gray]{0.9}  94.9\% & \cellcolor[gray]{0.9} 72.6\% \\
                                    &  Timing Slack &  -64.8\% &  3.4\% \\
           \multirow{2}{*}{a5/1}   & \cellcolor[gray]{0.9} Area         & \cellcolor[gray]{0.9} 69.8\% & \cellcolor[gray]{0.9} 69.4\% \\
                                   &  Timing Slack &  -27.2\% &  -1.2\% \\                        
           \hline \hline
        \end{tabular}
        
        \label{tab:over_javs}
    \end{table}
    
Utilizing exactly the same concepts and metrics described in \cite{Jagasivamani2014}, Otero \etal \cite{Otero2015} proposed a ``trusted split-foundry toolflow'' based on cellTK \cite{cellTK}. The concept of the  cellTK-based flow is to have on-demand cell generation from a transistor-level netlist. This is heavily in contrast with a traditional ASIC flow that relies on a predefined (and thoroughly characterized) cell library. Leveraging cellTK, \cite{Otero2015} proposed an extension referred as split-cellTK. This extension can generate multiple physical layouts for each unique cell without modifying the circuit topology, which is then used to implement obfuscation strategies. Two strategies are proposed, referred as Uniform and Random. The Uniform strategy tries to standardize the size and spacing between cells by inserting dummy transistors to equalize the number of nMOS and pMOS devices and, after the cell placement, dummy cells are inserted in empty gaps. A Random strategy is also proposed in order to reduce the overheads introduced by the Uniform strategy. Instead of deliberately standardizing the size and spacing between cells, a specific number of empty spaces is chosen for these tasks. A more in-depth explanation about their strategies is beyond the scope of this work because they are closely related to cellTK itself. However, their goals and evaluations are the same as in \cite{Jagasivamani2014}. For their experimental study, they utilized the island-style asynchronous FPGA developed in \cite{afpga}. A test chip was Split Manufactured in 65nm and the design was synthesized with a cellTK-based approach, i.e., without any defense strategy. Their defense strategies were evaluated only by simulation. The performance results for the baseline, Uniform, and Random strategies, when applied to obfuscate an adder, are shown in Table~\ref{tab:over_celltk}. Trustworthiness results are given in terms of neighbor connectedness and, for all implementations discussed, neighbor connectedness results were significantly smaller than the results reported in \cite{Jagasivamani2014}.
    
    \begin{table}[htb]
    \rowcolors{2}{gray!25}{white}
        \centering
         \caption{Performance Results Reported by \cite{Otero2015} to Obfuscate an Adder Circuit.}
        \begin{tabular}{lllll}
        \hline \hline \\ [-1.5ex]
         \parbox{1cm}{\textbf{Technique}} & \parbox{0.6cm}{\textbf{Area}\\ ($\mathbf{\mu m^2}$)} & \parbox{0.6cm}{\textbf{Power (mW)}} & \parbox{0.6cm}{\textbf{Energy (pJ)}} & \parbox{0.7cm}{\textbf{Perf. (MHz)}} \vspace{5pt} \\
          \hline
           Baseline  & 462 & 0.146 & 0.257 & 568 \\
           Uniform   & 717 & 0.149 & 0.307 & 486 \\
           Random    & 760 & 0.164 & 0.303 & 542 \\
           \hline
           \hline
        \end{tabular}
       
        \label{tab:over_celltk}
    \end{table}
  
In the context of obfuscation, but also generally for Split Manufacturing, a higher level of security is achieved when the chosen layer to perform the split is the lowest possible. Xiao \etal \cite{Xiao2015} pointed out that splitting at lower metal layers could increase the cost to manufacture the IC; it is argued that the FEOL-BEOL integration process must be more `precise' for correct alignment. Thus, a closer technology match between the trusted and untrusted foundries is required. As we previously argued, if the goal is to make use of the best silicon available from an untrusted foundry, the implication is that the trusted foundry cannot provide a legacy technology, but perhaps a mature yet still relevant technology is sufficient. In \cite{Xiao2015}, a methodology for obfuscating the layout is proposed for split at M3 or higher, meanwhile keeping the cost as low as possible and at the same time providing a high level of security. Their strategy is similar to the insertion of dummy cells; however, functional cells are inserted instead. Referred as obfuscated built-in self-authentication (OBISA) cells, the inserted functional cells are connected together to form a circuit. As the circuit is connected to the original circuit it is trying to protect, they claim this fact makes it extremely difficult for an attacker to separate the OBISA design from the original design. The idea behind OBISA is to obfuscate the layout by hindering neighbor connectedness analysis and standard-cell composition bias analysis while also perturbing the proximity between gates. As illustrated in Figure~\ref{fig:obisa}, two additional functional cells, O1 and O2, were placed between gates G2 and G3, and, G3 and G4, respectively. The insertion of these additional functional cells could deceive proximity attacks, assuming that the EDA tool would place the gates between OBISA cells farther apart than in the original circuit.
    
The proposed OBISA circuitry has two operating modes: functional and authentication. During functional mode, the OBISA circuitry stays idle and incoming signals and clock are gated/blocked. Thus, the original circuit is not affected by OBISA operating as it should. As the name suggests, when in authentication mode, OBISA is used to verify the trustworthiness of the manufactured IC (in the field). The specifics of the authentication are beyond the scope of this work and will not be discussed. The insertion of OBISA cells follows a similar strategy of dummy cell insertion as discussed in \cite{Jagasivamani2014}. The connections of the inserted cells are done in a way to promote the testability of the OBISA circuit and increase the obfuscation strength. Their approaches were evaluated using benchmark circuits from OpenCores. Results for the smallest and largest circuits are shown in Table~\ref{tab:obisa_res}. 
     
     \begin{table}[htb]
     \rowcolors{2}{gray!25}{white}
         \centering
         \caption{Implementation Results from \cite{Xiao2015}.}
         \begin{tabular}{lllll}
            \hline \hline \\ [-1.5ex]
              \textbf{Benchmark} &  \parbox{0.8cm}{\textbf{Gate count}} & \parbox{1.5cm}{\textbf{OBISA cell count}} & \textbf{Total nets} & \textbf{Nets} $\mathbf{\geq M4}$  \vspace{5pt}\\
              \hline
              DES3 & 1559 & 158 & 1799 & 127 \\
              DES\_perf & 49517 & 2090 & 49951 &  1343 \\
              \hline \hline
         \end{tabular}
         \label{tab:obisa_res}
     \end{table}
    
    \Figure[t!](topskip=0pt, botskip=0pt, midskip=0pt)[width=0.95\linewidth]{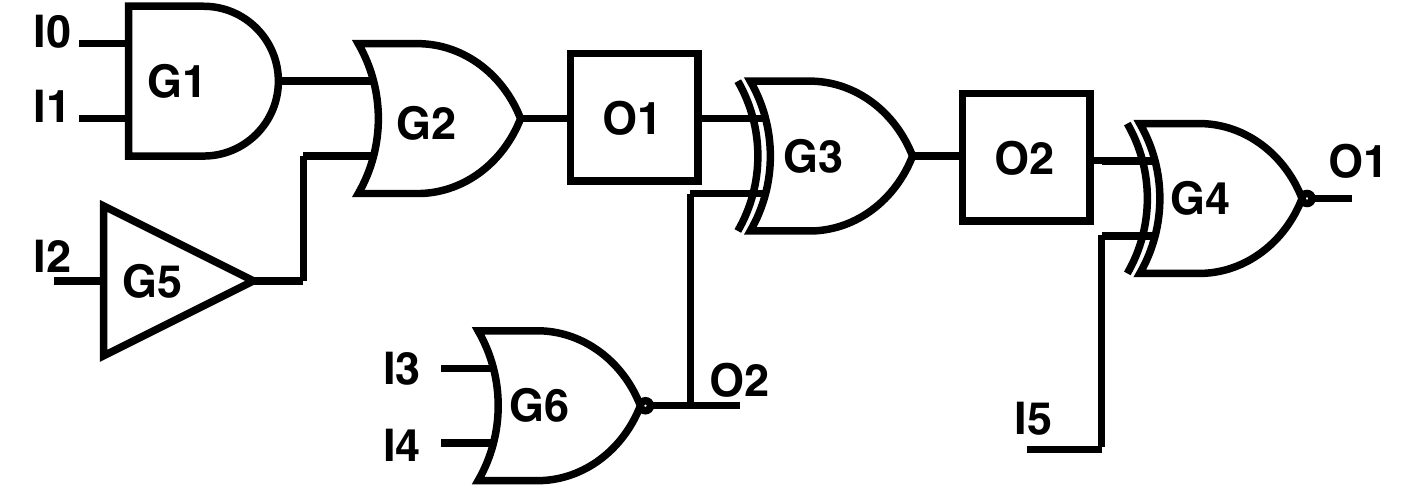}
{Circuit representation with OBISA cells (square cells) inserted (adapted from \cite{Xiao2015}). \label{fig:obisa}}
     
Another study using look-alike cells is reported by Masoud \etal \cite{Masoud2019} where the goal remains to make the attacker unable to distinguish cells and their inputs/outputs, thus mitigating attacks to some degree. In this study, two types of search algorithms are proposed to replace cells for isomorphic cells. In contrast with \cite{Jagasivamani2014} where all cells are replaced, in \cite{Masoud2019}, only cells with high impact on the security are replaced. Thus, the overhead introduced by cell replacement can be controlled (i.e., a trade-off is established). The proposed algorithms are based on `gate normalization', whereby truth tables of cells are analyzed in order to balance the occurrence of 0s and 1s (e.g., XOR and XNOR gates are normalized by definition). An analysis is made by replacing existing gates by XORs and comparing the deviation from the original circuit. If the deviation is larger than a given deviation threshold, the gates are effectively replaced.

A novel layout obfuscation framework is proposed by Li \etal \cite{Li2019a} which builds on the wire lifting concept of \cite{Imeson2013}. According to the authors, wire lifting alone is not enough to secure a design. If an attacker can tell the functionality of a specific gate that had its wires lifted, the security is already compromised. To address this problem, a framework that considers dummy cells and wire insertion simultaneously with wire lifting is proposed. As in \cite{Imeson2013}, threat model II was used. The proposed framework makes use of mixed-integer linear programming (MILP) formulation for the FEOL layer generation and a Lagragian relaxation (LR) algorithm to improve scalability. The generation of the new FEOL layout considers three operations: wire-lifting, dummy cell insertion, and dummy wire insertion. Dummy wire insertion is done only on dummy cells; thus, the original functionality of the circuit is guaranteed to remain and floating pins are avoided. Utilizing a graph representation, they re-formulate the security metric to accommodate dummy cell and dummy wire insertion. Since the original graph isomorphic relationship is lost when new nodes are inserted, a new approach has to be used to formalize the relationship between the original and the new FEOL; this concept is denoted as k-isomorphism \cite{KIsomorph} and the associated security analysis is denoted as \textit{k-security}. In their experimental study, TrustHub \cite{TrustHub} trojan insertion methods are used to select the nodes for protection. They used ISCAS\textquotesingle85 benchmark circuits together with functional units (shifter, alu, and div) from the OpenSPARC T1 processor \cite{openSPARK}. Comparison between MILP and LR algorithms are done for several \textit{k-security} levels, and the results are given in terms of area overhead (AO) and wire-length overhead (WLO). The results for a few of the security levels are shown in Table~\ref{tab:li_milplr}. 
     
     \begin{table}[htb]
     \rowcolors{2}{gray!25}{white}
         \centering
         \caption{Comparison Between MILP and LR Algorithms for the c4232 circuit \cite{Li2019a}.}
         \begin{tabular}{clll}
         \hline \hline \\ [-1.5ex]
             \textbf{Security Level } & \textbf{Algorithm} & \textbf{AO(\%)} & \textbf{WLO(\%)} \vspace{5pt}\\
             \hline
              15 & MILP & 18& 180\\
              20 & MILP & 41 & 220\\
              25 & MILP & 58 & 295\\
              15 & LR & 18& 200\\
              20 & LR & 40 & 230\\
              25 & LR & 60 & 305\\
            \hline \hline
         \end{tabular}
         
         \label{tab:li_milplr}
     \end{table}
     
\section{Future trends and challenges} \label{sec:trends}

Despite our effort to present the results of the many studied papers in the most fair way possible, it is clear that the hardware security community lacks a \textit{unified benchmark suite} and/or a \textit{common criteria} for assessing results. Often, researchers make use of benchmark suites that are popular in the Test community but have no real applicability in security. For instance, the ISCAS\textquotesingle85 suite has no crypto cores in it, which are the bread and butter of the research in the area. Furthermore, we believe the community would largely benefit from using circuits that better represent IC design practices of this decade where IPs often have millions of gates and ICs have billions of transistors.

While the lack of a common criteria is an issue for the academic community, the lack of an industry-supported path for Split Manufacturing is even more troubling. Today, more than ever, foundries compete for the title of `best silicon' and rarely engage in cross-foundry cooperation. Efforts of the past, such as the now defunct Common Platform of IBM, Samsung and GF, could have been a catalyzer for the adoption of Split Manufacturing. Without such collaboration, it is hard to foresee a future where the technique will gain traction again. Furthermore, the study of DFM-related implications of the technique is really cumbered by the fact that we cannot measure yield from massively produced Split Manufactured chips. 

We have discussed in details how many attacks leverage heuristics and hints left behind by the EDA tools. Many of these hints are very logical and can be appreciated, even graphically, as we illustrated in Figure~\ref{fig:prox_sa}. It is entirely possible that machine learning approaches can detect subtle biases in the tools that are not easy to appreciate graphically. There is no consolidated knowledge of what these biases are and to which extent machine learning is effective in detecting them. This avenue of research is certainly interesting and we believe it will be the target of many papers in the near future.

    \Figure[t!](topskip=0pt, botskip=0pt, midskip=0pt)[width=0.95\linewidth]{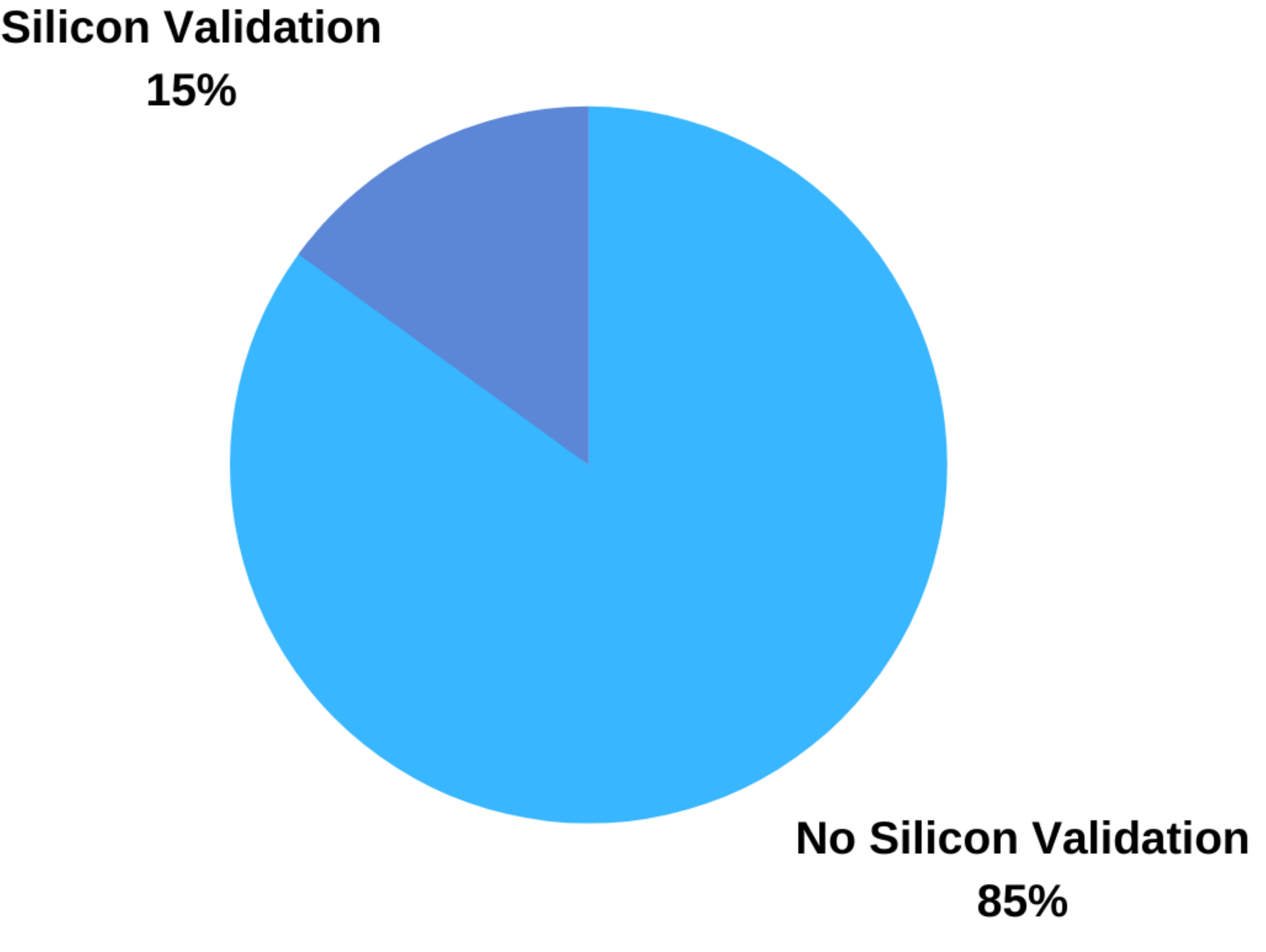}
{Techniques validated in silicon among presented works. \label{fig:pizza}}

It is also worth discussing the attack models that have been proposed so far. We have previously highlighted how strong Threat model II is, but the fact of the matter is that defining the threat model is a complicated exercise in which we seek to establish what are the capabilities of the attacker. By definition, formalizing the capabilities of an attacker requires understanding his motivations, technical proficiency, and availability of resources. If the attacker is underestimated, useless defense strategies can be devised and assumed to be effective. If the attacker is overestimated, convoluted defense strategies might be employed, leading to unnecessary PPA overheads. This is a challenge for Split Manufacturing and many other techniques that promote obfuscation.

Another topic that has led to no consensus is whether an attacker can make use of a partially recovered netlist. For instance, let us assume a design that instantiates the same block multiple times. If one of the blocks is correctly recovered, perhaps a cursory inspection of the structure will allow the attacker to recover all other instances of the same block. The same line of thinking can be applied to datapaths and some cryptographic structures that are regular in nature. In a sense, an analysis of the functionality of the recovered netlist could be combined with existing attacks for further improvement of correctly guessed connections.

We note that many of the works studied in this survey have not actually demonstrated their approach in silicon. This fact is summarized in Figure~\ref{fig:pizza}. As a community effort, we should strive to validate our approaches in silicon as often as possible. However, as discussed before, finding two foundries willing to diverge from their established practices could be next to impossible. This is likely the main reason that such small percentage of the works herein reported have validated their techniques in silicon. 

\section{Conclusion} \label{sec:conclusion}  
  
Our findings showed a big disparity on how the Split Manufacturing technique is approached among the surveyed studies. A variety of benchmark suites and metrics were used for evaluation, making direct comparisons between studies very difficult -- and, in some cases, impossible. In spite of that, we were able to classify the studies, clearly demonstrating the many interpretations of the technique, its attacks, and defenses. Our belief is that this survey assesses the most significant studies about Split Manufacturing as we focused on papers that appear on highly-regarded venues. Results gathered from the surveyed studies were compiled such that main features, metrics, and performance results are available. Regarding the results themselves, these are presented in such manner to illustrate the present state of the technique. Therefore, this work can be very helpful for future researchers to contextualize their own techniques for augmenting Split Manufacturing.
 
Overall, the security of Split Manufacturing is still under debate. Some studies conclude that the technique is indeed secure, and others that it is not. However, these conclusions are reached for different scenarios, i.e., using different benchmark circuits and set of metrics. Creating a unified benchmark suite suitable for Split Manufacturing evaluation, along with a unified set of metrics to quantify/qualify its performance, could facilitate the discussion about its security. In addition, increasing the number of demonstrations in silicon could also help with evaluation and adoption issues related to Split Manufacturing.

\bibliographystyle{ieeetr}
\bibliography{split_refs}

\begin{IEEEbiography}[{\includegraphics[width=1in,height=1.25in,clip,keepaspectratio]{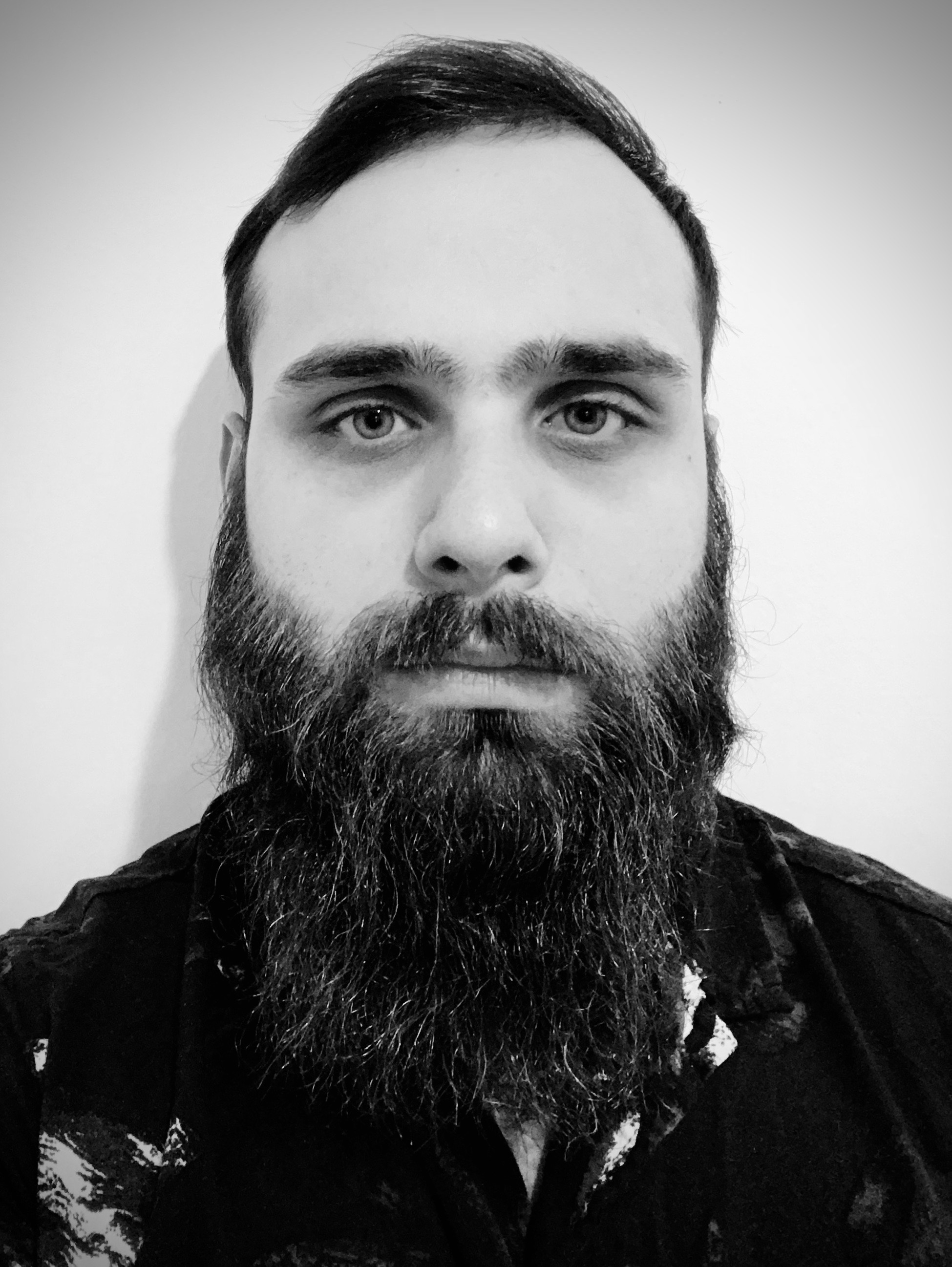}}]{Tiago D. Perez} received the M.S. degree in electric engineering from the University of Campinas, S\~ao Paulo, Brazil, in 2019. He is currently pursuing a Ph.D. degree at Tallinn University of Technology (TalTech), Tallinn, Estonia. 

From 2014 to 2019, he was a Digital Designer Engineer with Eldorado Research Institute, S\~ao Paulo, Brazil. His fields of work include digital signal processing, telecommunication systems and IC implementation. His current research interests include the study of hardware security from the point of view of digital circuit design and IC implementation.

\end{IEEEbiography}

\begin{IEEEbiography}[{\includegraphics[width=1in,height=1.25in,clip,keepaspectratio]{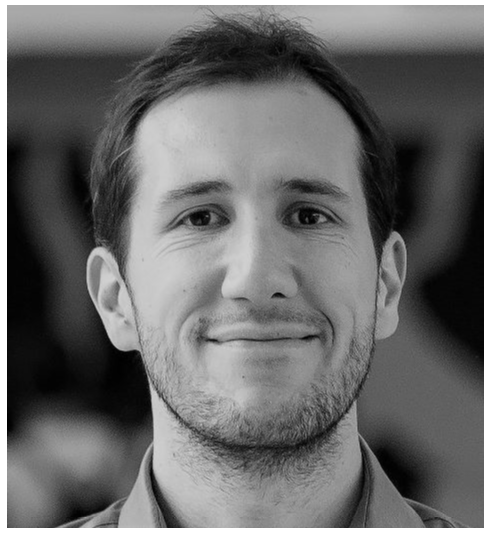}}]{Samuel Pagliarini}
(M'14) received the PhD degree from Telecom ParisTech, Paris, France, in 2013. 

He has held research positions with the University of Bristol, Bristol, UK, and with Carnegie Mellon University, Pittsburgh, PA, USA. He is currently a Professor of Hardware Security with Tallinn University of Technology (TalTech) in Tallinn, Estonia where he leads the Centre for Hardware Security. His current research interests include many facets of digital circuit design, with a focus on circuit reliability, dependability, and hardware trustworthiness.

\end{IEEEbiography}

\EOD

\end{document}